\documentclass[a4paper,fleqn,usenatbib]{mnras}

\usepackage[T1]{fontenc}
\usepackage{ae,aecompl}

\usepackage{graphicx}	% Including figure files
\usepackage{amsmath}	% Advanced maths commands
\usepackage{amssymb}	% Extra maths symbols

\title[Polarised radiative transfer around black holes]{A public code
  for general relativistic, polarised radiative transfer around
  spinning black holes}
\author[Dexter]{Jason Dexter$^{1}$\thanks{E-mail: 
jdexter@mpe.mpg.de}\\
$^{1}$Max Planck Institute for Extraterrestrial Physics,
Giessenbachstr. 1, 85748 Garching, Germany\\
}
\begin{document}
\pagerange{\pageref{firstpage}--\pageref{lastpage}} \pubyear{2015}
\maketitle

\label{firstpage}

\begin{abstract}
Ray tracing radiative transfer is a powerful method for comparing
theoretical models of black hole accretion flows and jets with
observations. We present a public code, \textsc{grtrans}, for
carrying out such calculations in the Kerr metric, including the full treatment of
polarised radiative transfer and parallel transport along
geodesics. The code is written in Fortran 90 and efficiently
parallelises with OpenMP, and the full code and several components
have Python interfaces. We describe several tests which 
are used for verifiying the code, and we compare the results for polarised
thin accretion disc and semi-analytic jet problems with those from the
literature as examples of its use. Along the way, we provide
accurate fitting functions for polarised synchrotron emission and transfer coefficients 
from thermal and power law distribution functions, and compare results
from numerical integration and quadrature solutions of the polarised radiative transfer
equations. We also show that all transfer coefficients can play an
important role in predicted images and polarisation maps of the
Galactic center black hole, Sgr A*, at submillimetre wavelengths.\end{abstract}


\section{Introduction}

Quantitative comparisons of theoretical models of black hole accretion
flows and jets with observations require radiative transfer calculations. The
bulk of the radiation is often produced near the black hole event
horizon, where relativistic effects of Doppler beaming, gravitational
redshift, and light bending become important. Ray tracing is a convenient
method for carrying out fully relativistic radiative transfer
calculations. Light bending is naturally accounted for by taking the
rays to be null geodesics in the Kerr metric, and the radiative transfer
equation can then be solved along geodesics to calculate observed
intensities. 

This technique has been used to calculate images \citep[e.g.,][]{luminet1979}
and spectra \citep[e.g.,][]{cunningham1975} of thin black hole accretion
discs \citep{shaksun1973,page1974}, including state of the art
methods to fit spectra in order to infer parameters such as the black
hole spin
\citep{davisetal2006,lietal2005,dauseretal2010}. Ray tracing is also convenient for including general
relativistic rotations of the polarisation direction via parallel
transport \citep{connorsstark1977,connorspiranstark1980}, and has been applied to the polarised radiative transfer of
synchrotron radiation from thick accretion discs, e.g. in order to 
model the Galactic center black hole Sgr A*
\citep{broderickloeb2005,broderickloeb2006}. With the development of
general relativistic MHD simulations of black hole accretion
\citep{devilliers2003,gammie2003}, ray tracing has become popular as a
post-processing step to study their variability properties 
\citep{schnittman2006,noblekrolik2009,dexterfragile2011} and radiative
efficiency \citep{nobleetal2011,kulkarnietal2011}, as well as for
comparison with observations of Sgr A*
\citep[e.g.,][]{noble2007,moscibrodzka2009,dexter2009,chanetal2015,goldmckinney2016}
and M87 \citep[e.g.,][]{dexteretal2012,moscibrodzkaetal2015}. 

Of particular interest are radiative transfer calculations relevant
for current and future event horizon scale interferometric observations of Sgr A* and M87
at submillimeter \citep[The Event Horizon Telescope,][]{doelemanetal2009} and near-infrared \citep[the VLTI GRAVITY
instrument,][]{eisenhauer2008} wavelengths. Fully modeling the observed
synchrotron radiation requires polarised
radiative transfer. Existing codes for this application are either
private \citep{broderickblandford2004} or written as post-processors
to specific numerical simulations \citep{shcherbakovetal2012}. Other
public tools \citep[e.g., \textsc{Gyoto} and
\textsc{Kertap},][]{vincentetal2011,chenetal2015} do not include fully polarised
radiative transfer.

We present a publicly available, fully general relativistic code,
\textsc{grtrans}\footnote{\url{https://www.github.com/jadexter/grtrans}},
for polarised radiative transfer via ray tracing in the Kerr metric. We describe the 
methods used for the parallel transport of the polarisation basis into the local frame of
the fluid (\S \ref{sec:parallel}) and the integration of the polarised radiative
transfer equations (\S \ref{sec:integration}) using emission,
absorption, and rotation coefficients (\S \ref{sec:transf-coeff}) calculated based on
radiative processes in terms of a background
fluid model (\S \ref{sec:fluid-models}). In \S
\ref{sec:tests-examples}, we discuss tests used to validate
the code, and comparisons of full example problems to those in the
literature. We also provide fitting functions for polarised
synchrotron emission, absorption, and transfer coefficients (Appendix
\ref{sec:farad-coeff-power} and \ref{sec:farad-coeff-power}), and show
an example polarised image from a model of the submm 
emission of Sgr A*, to demonstrate how all of the transfer
coefficients play important roles in the final polarised
image. Finally, \S \ref{sec:code-struct-perf} gives a summary of the
code convergence and performance properties, and an overview of its
organisation. 

\section{Methods}

The goal of a ray tracing radiative transfer code is to calculate the
observed intensity on locations (pixels) of an observer's
camera for a given model of emission and absorption. We calculate the
Boyer-Lindquist coordinates of the photon trajectories 
from the observer towards the black hole (trace the rays) corresponding to 
each pixel, parallel transport the observed polarisation basis into the fluid
frame, calculate the local emission and absorption properties at
each location, and then solve
the radiative transfer equations for the given emission and abosrption
along those rays.

\subsection{Ray tracing}

The observer's camera at inclination
$\mu_0 = \cos{\theta_0}$ and orientation $\phi_0$ has pixels whose coordinates are
described by apparent impact parameters $\alpha$, $\beta$ parallel and
perpendicular to the black hole spin axis. The photon trajectories in
\textsc{grtrans} are assumed to be geodesics in the Kerr metric, in
which case their constants of motion are specified for given
$\alpha$, $\beta$ \citep{bardeen1972}:

\begin{eqnarray}
l = -\alpha \sqrt{1-\mu_0^2},\\
q^2 = \beta^2 + \mu_0^2(\alpha^2-a^2),
\end{eqnarray}

\noindent where $l$, $q^2$, and $a$ are the dimensionless z-component of the
angular momentum, Carter's constant, and black hole spin parameters.

The trajectories for each ray given the constants
can then be found by solving the geodesic equation. We do this
semi-analytically by reducing the equations of motion to Jacobian
integrals and Jacobi-elliptic functions \citep{rauchblandford,agolphd}
as implemented in the code \textsc{geokerr} \citep{dexteragol2009}. 

In this method, the independent variable is either the inverse radius
$u = 1/r$ or $\mu = \cos{\theta}$. The former is used by default,
since even steps in $u$ naturally concentrate resolution towards the
black hole, where most of the radiation is produced. In
special cases, for example a thin accretion disc in the equatorial
plane, the latter method is preferable since then one can solve
for the radius where $\mu = 0$, without needing to
integrate the geodesic. In the default case with $u$ 
as the independent variable, the sampling can become poor near radial 
turning points (e.g. sections of the orbit at nearly constant radius). For this reason,
near radial turning points $\mu$ is instead used as the independent
variable to fill in the geodesic.

The calculation is started at a small, non-zero value of $u$ in order
to keep the coordinate time and affine parameter finite. The geodesics
are tabulated starting at a value of $u$ of interest for the problem
(e.g. the outer radial boundary of a numerical simulation) and are
terminated either just outside the event horizon for bound orbits, 
or once they again reach the outer radius of interest for the
calculation. The locations to sample ($u_i$) and number of samples ($n$) are code
parameters. The assumption made by the code is that the initial
intensity is zero at the farthest point sampled along the ray. 

\subsection{Parallel transport of the polarisation basis}
\label{sec:parallel}

The observed polarisation is measured with respect to the
horizontal and vertical axes defining the camera, while the polarised
emission and transfer coefficients are most naturally given relative
to a local direction in the emitting fluid (e.g. the magnetic field direction for
synchrotron radiation). To relate the two, we first parallel transport the observed
polarisation basis along the geodesic, and then transform it to the
orthonormal frame comoving with the fluid. The angle between the two
bases can then be used to rotate the local coefficients into the
observed polarisation basis. 

Parallel transport of a vector describing the polarisation basis
$f^\mu$ perpendicular to the wave-vector $k^\mu$ is simplified in the Kerr metric by the existence of 
of a complex constant called the Walker-Penrose constant
\citep{walkerpenrose1970}, given in
Boyer-Lindquist coordinates with $G=c=M=1$ as
\citep{connorsstark1977,connorspiranstark1980,chandrasekhar83}: 

\begin{eqnarray}\label{walkerpenrose}
&&K_1-i K_2 = (r-ia\cos{\theta})
\left\{(k^tf^r-k^rf^t)+a\sin^2{\theta}(k^rf^\phi-k^\phi f^r)\right.\nonumber\\
&&\left.-i[(r^2+f^2)(k^\phi f^\theta-f^\phi k^\theta)-a(k^tf^\theta-k^\theta f^t)]\sin{\theta}\right\},
\end{eqnarray}

\noindent where

\begin{eqnarray}
k^t &=& \frac{1}{\rho^2}\left[-a \left(a \sin^2{\theta} - l\right) +
  \frac{\left(r^2 + a^2\right)}{\Delta} \left(r^2+a^2-a l\right)\right],\\
k^r &=& -\frac{s_r}{\rho^2} R(r),\\
k^\theta &=& -\frac{s_\theta}{\rho^2} \sqrt{\frac{M(\theta)}{\sin^2{\theta}}},\\
k^\phi &=& \frac{1}{\rho^2} \left[-a +
  \frac{l}{\sin^2{\theta}}+\frac{a}{\Delta} (r^2+a^2-a l)\right],\\
M(\theta) &=& q^2 + \left(a^2-q^2-l^2\right) \cos^2{\theta} - a^2 \cos^4{\theta},\\
R(r) &=& r^2 + (a^2-q^2-l^2) + 2 \left[(a-l)^2+q^2\right] r^{-1}\\
&-&a^2 q^2 r^{-2},\\
\rho^2 &=& r^2+a^2 \cos^2{\theta},\\
\Delta &=& r^2 - 2 r + a^2,\\
\end{eqnarray}

\noindent is the photon wave vector whose direction is specified by
the signs $s_r$ and $s_\theta$ \citep[e.g.,][]{rauchblandford}.

The real and
imaginary parts of the constant, $K_1$ and $K_2$, provide two constraints on the
transported basis vectors, while the orthogonality condition $k^\mu
f_\mu=0$ provides a third. Since the polarisation basis vectors are
already only defined up to a multiple of the wave vector, we can set
$f^t = 0$ without any loss of generality, which leaves three
linear equations for the three remaining components of $f^\mu$:

\begin{eqnarray}
K_1 &=& \delta_1 f^r+\delta_2 f^\theta+\delta_3 f^\phi \\
K_2 &=& \gamma_1 f^r+\gamma_2 f^\theta+\gamma_3 f^\phi \\
k^\mu a_\mu&=&0
\end{eqnarray}

\noindent with

\begin{eqnarray}
\delta_1&=&rk^t-r a\sin^2{\theta}k^\phi \\
\delta_2&=&a^2\sin{\theta}\cos{\theta}k^t-a\cos{\theta}\sin{\theta}(r^2+a^2)k^\phi \\
\delta_3&=&ra\sin^2{\theta}k^r+a\cos{\theta}\sin{\theta}(r^2+a^2)k^\theta \\
\gamma_1&=& a\cos{\theta}k^t-a^2\cos{\theta}\sin^2{\theta}k^\phi\\
\gamma_2&=& r(r^2+a^2)\sin{\theta}k^\phi-ar\sin{\theta}k^t\\
\gamma_3&=& a^2\cos{\theta}\sin^2{\theta}k^r-r(r^2+a^2)\sin{\theta}k^\theta.
\end{eqnarray}

\noindent The components of $f^\mu$ can then be calculated as:

\begin{eqnarray}\label{partrans}
f^r&=&\frac{1}{N} \left[(\gamma_2 K_1 - \delta_2 K_2)(g_{\phi \phi} k^\phi + g_{\phi t}k^t) - \right.\\&&\left.(\gamma_3 K_1 - \delta_3 K_2) g_{\theta \theta} k^\theta \right],\\
f^\theta&=&-\frac{1}{N}\left[(\gamma_1 K_1 - \delta_1 K_2)(g_{\phi
    \phi} k^\phi + g_{\phi t} k^t) - \right.\\&&\left.(\gamma_3 K_1 - \delta_3 K_2) g_{r r} k^r  \right],\\
f^\phi&=&\frac{1}{N} \left[(\gamma_1 K_1 - \delta_1 K_2) g_{\theta \theta} k^\theta -\right.\\&&\left.(\gamma_2 K_1 - \delta_2 K_2)g_{r r} k^r \right],\\
N&=&(\gamma_2\delta_1-\gamma_1\delta_2)g_{\phi \phi} k^\phi - (\gamma_3\delta_1 - \gamma_1\delta_3)g_{\theta \theta} k^{\theta}\label{partrans2}\nonumber\\&+&(\gamma_2 \delta_1 - \gamma_1 \delta_2) g_{\phi t} k^t + (\gamma_3 \delta_2 - \gamma_2 \delta_2)g_{r r} k^r,
\end{eqnarray}

\noindent where $g_{\mu \nu}$ are the covariant metric components:

\begin{eqnarray}
g_{tt} &=& -\frac{1}{\rho^2 \Delta}\left[\left(r^2+a^2\right)^2-a^2
  \Delta \sin^2{\theta}\right],\\
g_{\phi t} &=& \frac{-2 a r \sin^2{\theta}}{\rho^2},\\
g_{r r} &=& \frac{\rho^2}{\Delta},\\
g_{\theta \theta} &=& \rho^2,\\
g_{\phi \phi} &=& \frac{\Sigma \sin^2{\theta}}{\rho^2},\\
\Sigma &=& (r^2+a^2)^2 - a^2 \Delta \sin^2{\theta}.
\end{eqnarray}

The polarisation basis at the camera is defined so that positive
Stokes Q is measured relative to the $\hat{\phi}_0$ axis. Transforming
the linear polarisation basis vectors $\hat{\phi}_0$ and
$\hat{\theta}_0$ at the camera then requires
knowledge of $K_1$ and $K_2$ for these vectors. These can be found from the asymptotic
form of equation \eqref{walkerpenrose} \citep{chandrasekhar83}. They are given by $K_1=-\gamma$,
$K_2=-\beta$ and $K_1=-\beta$, $K_2 = \gamma$ respectively, where
$\gamma=-\alpha-a\sin{\theta}_0$ \citep{connorspiranstark1980}. Then
equations (\ref{partrans}-\ref{partrans2}) allow us to calculate the polarisation
basis vectors at any point along the ray.

\subsubsection{Transformation to the orthonormal fluid frame}\label{sec:polrefvector}

The emission coefficients and the transfer matrix computed in the
fluid frame are defined in a basis aligned with a local reference
vector. For the case of synchrotron emission, it is convenient to use
the local magnetic field direction and so we use $b^\mu$ as this
vector in \textsc{grtrans} without loss of generality. In the case of
electron scattering in a thin accretion disc, the polarisation is
given relative to the disc normal vector, and so we assign the
variable $b^\mu$ to that vector.

Before integrating the radiative transfer equations, these
coefficients must be transformed to the observed polarisation
basis. This transformation requires finding the angle between the 
transported polarisation basis vectors and the polarisation reference
vector \citep{shcherbakovhuang2011}. We transform into the orthonormal frame comoving with the fluid where the
four-velocity is $\hat{u}^\mu=(1,0,0,0)$. The basis four-vectors of 
the transformation are 
\citep{kroliketal2005,beckwith2008,shcherbakovhuang2011,kulkarnietal2011}: 

\begin{eqnarray}
   e^\mu_{(t)} &=& u^\mu,\\
   e^\mu_{(r)} &=& (u_ru^t, -(u_t u^t+u_\phi u^\phi), 0, u_r
   u^\phi)/N_r,\\
   e^\mu_{(\theta)} &=& (u_\theta u^t, u_\theta u^r, 1+u_\theta u^\theta,
   u_\theta u^\phi)/N_\theta,\\
   e^\mu_{(\phi)} &=& (u_\phi, 0, 0, -u_t)/N_\phi,
\end{eqnarray}

\noindent where the upper (lower) indices are lowered (raised) with the Kerr
(Minkowski) metric and,

\begin{eqnarray}
   N^2_r &=& -g_{rr}(u_tu^t+u_\phi u^\phi)(1+u_\theta u^\theta),\\
   N^2_\theta &=& g_{\theta\theta}(1+u_\theta u^\theta),\\
   N^2_\phi &=& -(u_t u^t + u_\phi u^\phi)\Delta \sin^2{\theta},\\
   \Delta &=& r^2-2r+a^2.
\end{eqnarray}

\noindent Four-vectors in the coordinate frame are transformed as,

\begin{equation}
 A_{(\alpha)}=e^\mu_{(\alpha)} A_\mu.
\end{equation}

The angle $\chi$ between the projected magnetic field and the
polarisation basis is given in terms of ordinary dot products of the 
magnetic field and parallel-transported basis three-vectors (denoted by hats):

\begin{eqnarray}\label{chieq}
 \sin{2 \chi} &=& -2\frac{\left(\hat{a} \cdot
     \hat{B}\right)\left(\hat{b} \cdot \hat{B}\right)}{\left(\hat{a} \cdot
       \hat{B}\right)^2+\left(\hat{b} \cdot \hat{B}\right)^2},\\
 \cos{2 \chi} &=& \frac{\left(\hat{b} \cdot
       \hat{B}\right)^2-\left(\hat{a} \cdot \hat{B}\right)^2}{\left(\hat{a} \cdot
       \hat{B}\right)^2+\left(\hat{b} \cdot \hat{B}\right)^2}.
\end{eqnarray}

\noindent In this frame, the combined redshift and Doppler factor $g \equiv
\nu_0 / \nu = -1/\hat{k}^t$ and $\cos \theta_B={\frac{\hat{k}\cdot
    \hat{B}}{|\hat{k}||\hat{B}|}}$.

\subsubsection{Transfer Equation}

The non-relativistic polarised radiative transfer equation can be written in the form,

\begin{equation}\label{transfer}
\frac{d}{ds}
\left(\begin{array}{c}  I \\  Q \\  U \\  V \\\end{array}\right)
=
\left(\begin{array}{c}
  j_I \\  j_Q \\  j_U \\  j_V
\end{array}\right)-
\left(%
\begin{array}{cccc}
  \alpha_I & \alpha_Q & \alpha_U & \alpha_V \\
  \alpha_Q & \alpha_I & \rho_V & \rho_U \\
  \alpha_U & -\rho_V & \alpha_I & \rho_Q \\
  \alpha_V & -\rho_U & -\rho_Q & \alpha_I \\
\end{array}%
\right)
\left(\begin{array}{c}  I \\  Q \\  U \\  V \\\end{array}\right)
\end{equation}

\noindent where ($I$, $Q$, $U$,
$V$) are the Stokes parameters, $j_{I,Q,U,V}$ are the polarised
emissivities, $\alpha_{I,Q,U,V}$ are the absorption coefficients, and
$\rho_{Q,U,V}$ are the Faraday rotation and conversion coefficients. 

In the context of synchrotron radiation, the transfer equation can be
simplified by aligning the magnetic field with Stokes $U$, so that
$j_U=\alpha_U=\rho_U=0$. Then $j_Q$, $\alpha_Q$ ($j_V$, $\alpha_V$) correspond to the
emission and absorption coefficients for linear (circular)
polarisation and $j_I$, $\alpha_I$ are the unpolarised coefficients. The transfer coefficients $\rho_{Q,V}$
describe the effects of Faraday conversion and rotation respectively.

All coefficients are computed in the fluid
rest frame, where $\nu$ is the emitted frequency, related to the
observed frequency through $g$. Then the
transfer equation is recast into invariant form: 
$\mathcal{I}=g^3\mathbf{I}$,
$\mathcal{J}=g^2\mathbf{j}$, and 
$\mathcal{K}=g^{-1} \mathbf{K}$, where $\mathbf{I}$, $\mathbf{j}$ and
$\mathbf{K}$ are the intensity and emissivity vectors and the
transfer matrix from equation \eqref{transfer}.

Finally, we use the angle $\chi$ to rotate the emissivity and
absorption matrix in the fluid frame into that of the observer, such
that the radiative transfer equation becomes,

\begin{equation}
\label{genrelpolradtrans}
\frac{d\mathcal{I}}{d\lambda}= \hat{\mathcal{J}}-\hat{\mathcal{K}}\mathcal{I},
\end{equation}

\noindent where $\lambda$ is an affine parameter,
$\hat{\mathcal{J}}=g^2 R(\chi) \mathbf{j}$, $\hat{\mathcal{K}}=g^{-1}
R(\chi) \mathbf{K} R(-\chi)$, and 

\begin{equation}
 R(\chi) = \left(\begin{array}{cccc}
  1 & 0 & 0 & 0 \\
  0 & \cos{2\chi} & -\sin{2\chi} & 0 \\
  0 & \sin{2\chi} & \cos{2\chi} & 0 \\
  0 & 0 & 0 & 1 \\
\end{array}\right).
\end{equation}

\noindent This rotation transforms the fluid frame polarisation
basis to that at infinity, including the parallel transport of the
polarisation four-vector along the ray. 

Equation \ref{genrelpolradtrans} includes all relativistic
effects. The bending of light is accounted for by the calculation of
null geodesics \citep{dexteragol2009}, the
gravitational redshifts and Doppler shifts due to fluid motions are
included in $g$. 

This method, developed by \citet{shcherbakovhuang2011}, parallel
transports the polarisation basis along the ray and into the fluid
frame. This is similar to the approach of 
\citet{connorspiranstark1980}, who transported a local polarisation vector
$f^\mu$ from the fluid to the observer. \citet{gammieleung2012} derived a general formalism for
covariant polarised radiative transfer, and mathematically
showed the equivalence of the approach used here and alternative methods used by \citet{broderickblandford2004}
and \citet{schnittmankrolik2013}. We show tests and example problems
comparing results from these methods in \S \ref{sec:test-problems}.

\subsection{Transfer coefficients}
\label{sec:transf-coeff}

The transfer coefficients in equation \eqref{transfer} depend in general on the
physical properties of the radiating particles. Here we focus
specifically on the case of synchrotron radiation appropriate for
studying accretion flows at the lowest observed luminosities (e.g., Sgr
A*). Adding different emissivities such as bremsstrahlung to
\textsc{grtrans} would require a straightforward modification of the
code. 

In addition, the form of the transfer coefficients depends on the
underlying electron distribution function. The appropriate forms for
thermal and power law distributions are implemented in the
code. More general distribution functions can be built by combining
these components (e.g. a thermal distribution with a power law
``tail'' or a superposition of thermal distributions, Mao et al. in
prep.). In these special cases, the integral over the distribution
function can be analytically approximated to high accuracy in the
ultra-relativistic synchrotron limit \citep[e.g.,][]{maha}. The full
forms for the transfer coefficients as used in \textsc{grtrans} and
some of their derivations are given in Appendix 
\ref{cha:polar-synchr-emiss}. 

In addition to synchrotron coefficients, for test problems with
optically thick accretion discs the code uses (color-corrected) blackbody intensity
functions for the disc surface brightness. 

\subsection{Fluid models}
\label{sec:fluid-models}

The calculation of transfer coefficients for a particular emission
model requires knowledge of the fluid state variables of
spacetime coordinates. Depending on the model used, this can include
the electron density, magnetic field strength and orientation, and the
internal energy density in electrons. In \textsc{grtrans} we 
implement several fluid models from the literature. They are described briefly below, and used as code
examples and tests in \S \ref{sec:test-problems}. 

\subsubsection{Thin accretion discs}

The relativistic version \citep{page1974} of the standard thin disc
solution \citep{shaksun1973} for axisymmetric, steady accretion in the
equatorial plane is implemented and intended for use with a 
model for the emergent intensity from the disc (e.g., a blackbody). In
this case, the net polarisation is taken to follow the result of
electron scattering in a semi-infinite atmosphere
\citep{sobolev1963,chandrasekhar1950} specified relative to the disc
normal vector (\S \ref{sec:parall-transp-tests}).

\subsubsection{Alternative thin accretion discs}

We have also implemented a numerical version of the thin accretion
disc problem which inputs a temperature distribution $T(r,\phi)$ in
the equatorial plane. One example use of this is for calculating
spectra of inhomogeneous (or ``patchy") accretion discs
\citep{dexteragol2011}, as used for the polarisation calculations
described in \citet{dexterquataert2012}.

\subsubsection{Spherical accretion flow}

A solution of the general relativistic fluid equations for spherically
symmetric inflow in the Schwarzschild metric following
\citep{michel,shap1} is implemented. The dominant emission in this case comes from
synchrotron radiation \citep{shap2}. See \citet{dexteragol2009} for details. 

For polarised emission, we take
the magnetic field to be purely radial. This is done to check that the resulting linear polarisation sums to
zero (since for a camera centered on the black hole there is no
preferred direction), and the residual is used as an estimate of the minimum
systematic uncertainty in the fractional linear polarisation ($\simeq 0.01\%$). 

\subsubsection{Semi-analytic jet model}
\label{bl09jet}

\citet{broderickloeb2009} presented a semi-analytic jet model based on
stream functions found in force-free simulations. We have implemented this solution numerically on
a grid of ($r$,$\theta$) in Boyer-Lindquist coordinates. To generate
our numerical solutions, we solve for
the magnetic field and velocity structure analytically using their
equations 5-13. To get the particle density, we tabulate their
function $F(\psi)$ numerically using a separate grid of points with roughly
constant $z \simeq r_{\rm fp}$ and varying $\psi = r^{2-2\xi} (1 -
\cos{\theta})$. This function can then be used to calculate the particle
density (their equation 13). 

The fluid variable solutions from our method appear identical to what is shown in
their Figure 4.

\subsubsection{Numerical general relativistic MHD solution}

We also use another numerical solution, from the public version of the
axisymmetric general relativistic
MHD code \textsc{HARM} \citep{gammie2003,noble2006}. Starting from a gas torus
in hydrostatic equlibrium threaded with a weak magnetic field, the code evolves the equations of ideal MHD
in the Kerr spacetime. The magnetorotational instability \citep{mri}
drives turbulence in the torus and the resulting stresses transport
angular momentum outwards, leading to accretion onto the central black
hole. Snapshots from these simulations have been used as models of
Sgr A* \citep[e.g.,][]{noble2007,moscibrodzka2009}. 

The images used here as examples are from a single snapshot of a
simulation with black hole spin $a = 0.9375$ at $t = 2000 GM/c^3$ used for comparison with
3D simulations in \citet{dexteretal2010}. \textsc{grtrans} supports
fully time-dependent calculations using a series of such simulation
snapshots to e.g. calculate accretion flow movies rather than
images. It would also be straightforward to adapt the code to work
with updated \textsc{HARM} versions, for example with 3D data or
non-ideal MHD.

The fluid variables in these simulations are saved in modified Kerr-Schild
coordinates and with arbitrary units which assume
$G=c=M=1$. Calculating radiation from these data in \textsc{grtrans}
requires converting to Boyer-Lindquist coordinates and to cgs
units. The coordinate conversion is done analytically in two steps:
from modified to standard Kerr-Schild coordinates \citep{gammie2003}
and then from Kerr-Schild to Boyer-Lindquist coordinates 
\citep[e.g.,][]{font1999}. Scaling to cgs units is done by i) fixing the black
hole mass, which sets the length- and time-scales, and ii) choosing an
average accretion rate (or equivalently mass of the initial
torus). This procedure is discussed in more detail elsewhere 
\citep{schnittman2006,noble2007,dexteretal2010}.

Once the unit and coordinate conversions are done, we calculate fluid
variables at tabulated geodesic coordinates. For all numerical models,
we linearly interpolate from the set of nearest
neighbors on the grid for the numerical model. The way this is
implemented in the code assumes that the grid is uniformly spaced in
some coordinates, and the fluid model must include the transformation from those coordinates to 
Boyer-Lindquist. 

There are several other models implemented in the code in some form,
but which have not been tested. It is
straightforward to add new fluid models to the code, e.g. by using
existing ones as templates.

\subsection{Integration of the polarised radiative transfer equations}
\label{sec:integration}

From the previous steps, we have transfer coefficients specified at
tabulated points along a geodesic which are transformed to relativistic invariant form and
aligned with the observed Stokes parameters of the distant observer,
accounting for parallel transport along each ray. 

The final step is to solve the polarised radiative transfer (equation \ref{genrelpolradtrans}) along the
geodesic. In \textsc{grtrans}, this is done as a separate step
following the calculation of the coordinates of the geodesic. While
the ray tracing proceeds backwards from the camera towards the black
hole, the integration proceeds outwards. This is done so that we may safely 
set the initial intensity to zero at some point either where the
optical depth is large, or where the geodesic has left the emitting
volume. Here we describe one numerical
integration method and two quadrature methods that are implemented in
\textsc{grtrans} for integrating the equations.

These methods can be used for relativistic or non-relativistic
problems. For consistency with previous literature, we write the
non-relativistic versions of the intensity, absorption matrix, 
emissivity, and step size along the ray at index $k$ as $I_k$,
$K_k$, $j_k$ and $\Delta s_k$ in what follows. In \textsc{grtrans}, the
relativistic invariants $\mathcal{I}(\lambda_k)$,
$\hat{\mathcal{K}}(\lambda_k)$, $\hat{\mathcal{J}}(\lambda_k)$, and
$\Delta \lambda_k$ take the place of these quantities.

\subsubsection{Numerical integration}

The most straightforward method is numerical
integration of the radiative transfer equation. The
radiative transfer equations can be stiff: the required 
step size for a converged solution decreases sharply once $\tau
\gtrsim 1$, where $\tau$ is the optical depth associated with any
transfer coefficient. 

In order to get a robust solution, we use the
\textsc{ODEPACK} routine \textsc{LSODA} \citep{odepack} to advance the
Stokes intensities between each step tabulated on the geodesic. This
algorithm adaptively switches between a predictor-corrector (Adams) method for
non-stiff systems, and a BDF method for stiff systems. We find it
necessary to restrict the maximum step size allowed in $\lambda$,
since otherwise a large step can miss the region of interest
altogether. 

Regions of large optical depth often contribute negligibly to the
observed intensity but require a small step size, and
so we terminate the integration at a maximum optical depth,
$\tau_{max} = 10$ by default. There are further free
parameters in \textsc{LSODA} related to the error tolerance. 

The locations sampled by \textsc{LSODA} do not
correspond exactly to the points tabulated along the ray. We linearly interpolate the transfer coefficients
between tabulated points, even though they are highly non-linear
functions of position along the geodesic. The fluid variables vary
more smoothly along the ray, and it would be straightforward but more 
computationally expensive to instead re-interpolate the fluid
variables to the points used by \textsc{LSODA} and then calculate new
transfer coefficients, as was done in the previous version of the code
\citep{dexter2011}. Given the results from comparing to quadrature
integration methods and analytic solutions described below, and from
the convergence properties with increasing the number of points along
each ray, we find the current approximation adequate for obtaining
accurate solutions.

\subsubsection{Quadrature solutions}

The polarised radiative transfer equations are linear and ordinary,
and so admit a formal solution analagous to that of the unpolarised
case \citep{ryblight}. The solution amounts to finding the matrix operator
$\mathbf{O}(s,s')$, defined by \citep{deglinnocenti1985},

\begin{eqnarray}
\frac{d}{ds} \mathbf{O}(s,s') &=& -\mathbf{K}(s) \mathbf{O}(s,s'),\\
\mathbf{O}(s,s) &=& \mathbf{1},
\end{eqnarray}

\noindent which determines how the intensity is propagated over some
part of the ray in the absence of emission. In the unpolarised case this is a scalar, $O =
\exp{[-(\tau(s)-\tau(s')]}$. If the absorption matrix $\mathbf{K}$ is a constant over the ray, then
similarly,

\begin{equation}
\mathbf{O}(s,s') = \exp{[-\mathbf{K}(s-s')]}.
\end{equation}

\noindent In terms of $\mathbf{O}$, the intensity can be written in
terms of an initial value $I_{n-1} (s_{n-1})$:

\begin{equation}\label{eq:3}
\mathbf{I}(s) = \int_{s_{n-1}}^s ds' \mathbf{O}(s,s') j(s') + \mathbf{O}(s,s_{n-1}) I_{n-1}.
\end{equation}

\citet{deglinnocenti1985} found a closed form solution for
$\mathbf{O}$ (their equation 10, reproduced in Appendix
\ref{deglinnocentio}). This solution is valid for regions where the
transfer matrix $\mathbf{K}$ is constant, but not for our situation of
interest where they vary arbitrarily along a ray. In order to use this
expression, we assume that the coefficients are constant in between
the tabulated locations along a geodesic starting from $s_{k=n}$ at the
farthest point of interest along the ray (at the black hole or where
the ray leaves the far end of the emitting region) and integrating
towards $s_{k=0}$ (the ``surface''), and write the solution of 
equation \eqref{eq:3} separately for the interval between neighboring
points with indices $k+1$ and $k$ with positions $s_{k+1}$ and $s_k$:

\begin{equation}\label{omatrixmethod}
I_k = O_{k,k+1} j_k \Delta s_k + O_{k,k+1} I_{k+1},
\end{equation}

\noindent where $\Delta s_k = s_{k+1} - s_k$. This formula is used recursively going outwards from $s_{n-1}$
to $s_0$ to find the intensity everywhere from the initial condition
$I_{n-1}=0$. 

The final integration method implemented in \textsc{grtrans} is the
diagonal element lambda operator method \citep[DELO,][]{reesetal1989},
which comes from writing the transfer equations in terms of the unpolarised
optical depth, $d\tau = \alpha_I ds$, and the modified absorption matrix $K' =
\mathbf{K}/\alpha_I - \mathbf{1}$ and source function $S' = \mathbf{j} / \alpha_I$: 

\begin{equation}
\frac{d\mathbf{I}}{d\tau}=\mathbf{I}-\mathcal{S},
\end{equation}

\noindent where $\mathcal{S} = S' - K' \mathbf{I}$. This equation has
a formal solution between neighbouring points $\tau_{k+1}$, $\tau_{k}$
of 

\begin{equation}\label{deloeq}
I(\tau_{k}) = E_k I(\tau_{k+1}) + \int_{\tau_k}^{\tau_{k+1}} \exp[-(\tau-\tau_k)] \mathcal{S} d\tau,
\end{equation}

\noindent where $E_k = \exp(-\delta_k)$ and $\delta_k =
\tau_{k+1}-\tau_k$. The DELO method makes a linear approximation for
the modified source function, 

\begin{equation}
\mathcal{S}(\tau) = [(\tau_{k+1}-\tau)\mathcal{S}_k + (\tau-\tau_k)\mathcal{S}_{k+1}]/\delta_k,
\end{equation}

\noindent so that equation \eqref{deloeq} can be integrated analytically between grid points, giving:

\begin{equation}
I(\tau_k) = \mathcal{P}_k + \mathcal{Q}_k I(\tau_{k+1}),
\end{equation}

\noindent where 

\begin{eqnarray}
\mathcal{P}_k &=&  \mathcal{M}_k [(F_k-G_k)S'_k + G_k S'_{k+1}],\\
\mathcal{Q}_k &=& \mathcal{M}_k (E_k \mathbf{1}-G_k
K'_{k+1}),\\
\mathcal{M}_k &=& [\mathbf{1}+(F_k-G_k)K'_k]^{-1},\\
F_k &=& 1-E_k,\\
G_k &=& \left[1-\left(1+\delta_k\right)E_k\right]/\delta_k,
\end{eqnarray}

The difficulty with this method is that $\tau_k$ is used as the
independent variable. For our problems of interest $\tau$ 
can be nearly constant between grid points over which the fluid quantities and
emissivity change significantly, which causes the above solution to
fail. In the limit of small $\delta_k$, we instead expand the above
quantities up to $\mathcal{O}(\delta_k^2)$, leading to the following forms:

\begin{eqnarray}
\mathcal{P}_k &=& \mathcal{M}_k \left[\frac{j_k \Delta s_k}{2}-\frac{\Delta s_k^2 \alpha_{I,k}
    j_k}{6}+\frac{j_{k+1} \Delta s_k}{2}\right.\\&-&\left.\frac{\Delta s_k^2
    \alpha_{I,k} j_{k+1}}{3}\right],\\
\mathcal{Q}_k &=& \mathcal{M}_k
\left[\mathbf{1}\left(1-\frac{\Delta s_k \alpha_{I,k}}{2}+\frac{\Delta s_k^2
      \alpha_{I, k+1}^2}{6}\right)\right.\\&-&\left.\left(\frac{\Delta s_k}{2}-\frac{\Delta
      s_k^2}{3}\right)K_{k+1}\right],\\
\mathcal{M}_k &=& \left[\left(1-\frac{\delta_k}{2}+\frac{\delta_k^2}{6}\right)\mathbf{1}\right]^{-1}.
\end{eqnarray}

This version of the equations uses $s$ as the independent variable,
and is used by default when $\delta_k <
10^{-2}$. Since the number of steps 
taken by \textsc{grtrans} is usually $\gtrsim 400$, this form of the
equations is used unless the optical depth is very large.

In \textsc{grtrans}, all integration methods proceed outwards from an
initial point back towards the camera. From the recursive forms of the DELO and formal solution
methods, we see that it would also be possible to integrate the
polarised radiative transfer equations backwards by summing  
the so-called contribution vectors from each point to the final
intensity on the camera, $I_0$:

\begin{equation}
I_0 = \Sigma_{i=0}^{n-1} \mathcal{C}_i,
\end{equation}

\noindent where

\begin{equation}
\mathcal{C}_i = \left[\Pi_{m=0}^{i-1} O_{m,m+1}\right] j_i
\end{equation}

\noindent for the formal solution method and 

\begin{equation}
\mathcal{C}_i = \left[\Pi_{m=0}^{i-1} \mathcal{Q}_m\right] j_i
\end{equation}

\noindent for the DELO method \citep{reesetal1989}. The equivalent
contribution vectors in the unpolarised case are given as 

\begin{equation}
\mathcal{C}_i = \left[\Pi_{m=0}^{i-1} e^{-(\tau_{m+1}-\tau_m)}\right] j_i = e^{-\tau_i} j_i,
\end{equation}

\noindent where $\tau_i$ is the optical depth from the surface to the
depth at index $i$.

Tracing backwards from the camera, at each step at index $i$ one can
calculate $\mathcal{C}_i$ using the solution for $\mathcal{C}_{i-1}$,
$K_i$, and $j_i$. Solving the polarised radiative transfer equations
in this way would be useful in implementations where the geodesic
equations and radiative transfer equations are solved simultaneously,
e.g. as is done for the unpolarised case in the public code
\textsc{Gyoto} \citep{vincentetal2011}. In this method, one can then
safely terminate the integration early if the product term in
$\mathcal{C}_i$ becomes sufficiently small (e.g., the optical depth
becomes large).

The three methods give consistent answers, usually
to high accuracy and with similar performance. The main drawback of
our quadrature implementations is the lack of an adaptive step size,
so that many steps ($\sim 10^3$) are required in order to get a converged
result. In most example problems in the
following section, the numerical integrator is used as it is the most
robust choice. The other methods are primarily used for comparison and
testing, although they are faster at a fixed number of points $n$ and
so with some optimisation might prove to be significantly faster than
numerical integration.

\section{Tests and examples}
\label{sec:tests-examples}

Here we describe tests of the different aspects of \textsc{grtrans}
(unit tests), as well as full example problems which are compared with
results from the literature. We do not provide tests of the
\textsc{geokerr} code for calculating null geodesics in the Kerr
metric, which are described in \citet{dexteragol2009}.

\subsection{Parallel transport tests}
\label{sec:parall-transp-tests}

\begin{figure*}
\begin{center}
\begin{tabular}{cc}
\includegraphics[scale=0.62]{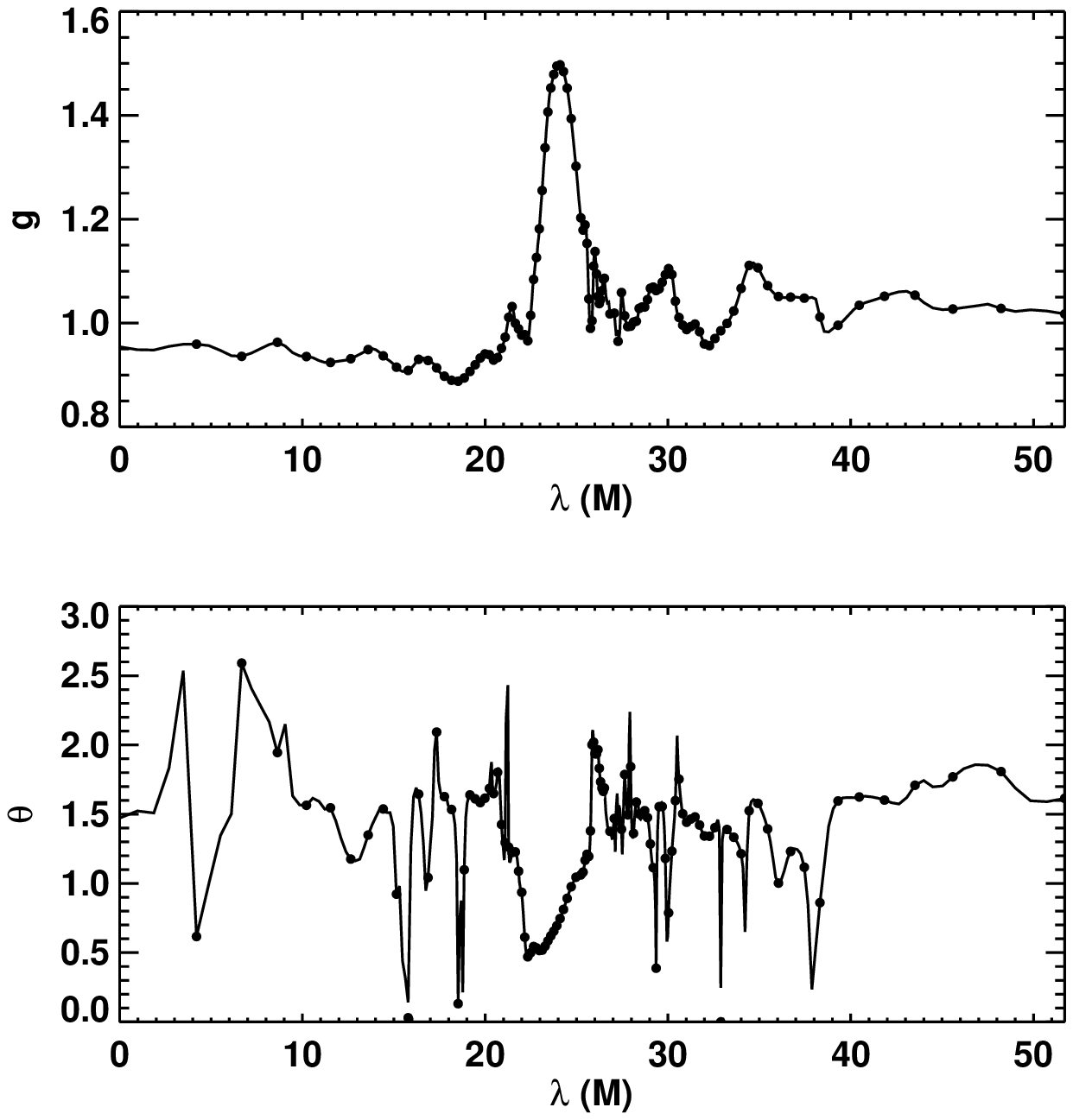}&
\includegraphics[scale=0.62]{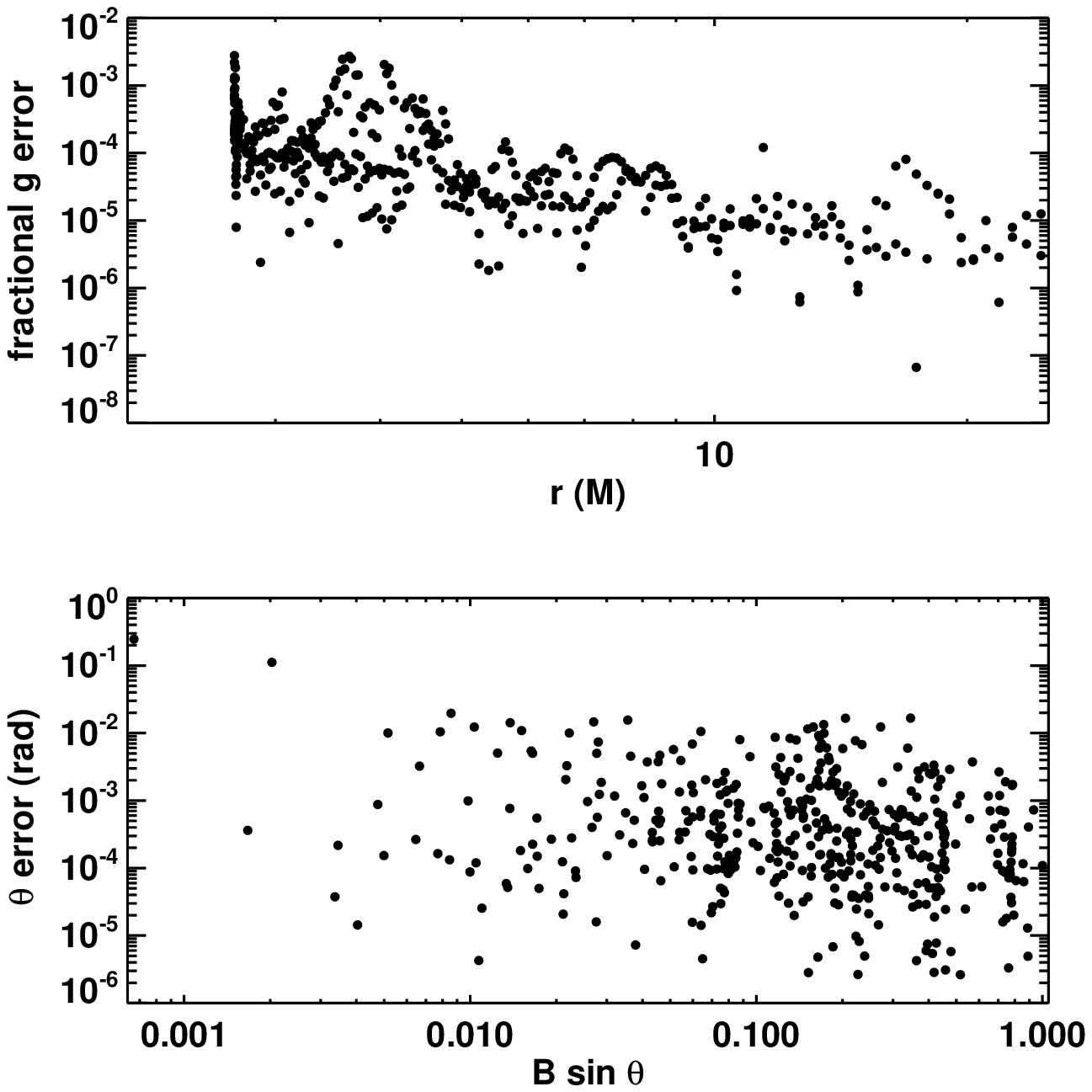}
\end{tabular}
\caption{\label{testrshiftang} Comparison between redshift/Doppler factor
  $g$ from \citet{viergutz1993} and angle between $k^\mu$ and $b^\mu$
  from \citet{broderick2004} in the fluid rest frame and those
  calculated 
  from transforming to the orthonormal fluid frame \citep[\S
  \ref{sec:parallel}][]{shcherbakovhuang2011} for a sample geodesic. The residuals give an idea for the systematic
  errors in these quantities, usually $< 1\%$. The error in $\theta_B$ can be
large when $k^\mu b_\mu$ is very small. However, for synchrotron radiation the
effect is negligible since here the emissivity is also small. For
clarity, only one of every 4 points is plotted.}
\end{center}
\end{figure*}

\begin{figure*}
\begin{center}
\begin{tabular}{cc}
\includegraphics[scale=0.5]{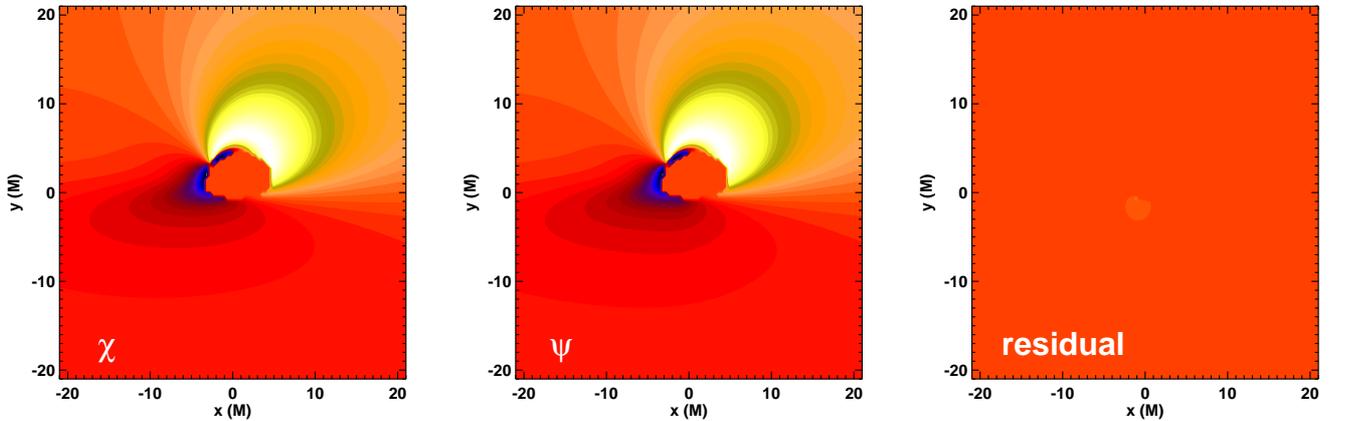}
\end{tabular}
\caption{\label{psichi}The rotation element $\sin 2\chi$ relating the
  Stokes parameters measured at the camera to those locally measured
  relative to lie in the plane of the disc. The left panel shows this quantity calculated by parallel
propagation of the camera back to the disc \citep[][and \S
\ref{sec:parallel}]{shcherbakovhuang2011} while the right panel is calculated by
parallel transporting the local polarisation vector to the camera
\citep{connorspiranstark1980,agolphd}. There is generally good agreement between the two
methods, although with up to $\simeq 10\%$ residuals near
the black hole in high spin cases.}
\end{center}
\end{figure*}

The accuracy of the method for the parallel transport of a vector in
the Kerr metric can be checked by calculating the Penrose-Walker
constant at each point along the ray, compared to the value at the
camera. In \textsc{grtrans}, this value remains constant along the ray
to machine precision. This result is expected, since the
parallel transport in the Kerr metric is done analytically (\S
\ref{sec:parallel}). 

The transported polarisation basis is compared to the polarisation
basis of the emission at each point in the so-called comoving
orthonormal frame \citep{shcherbakovhuang2011}, where the fluid
four-velocity $u^\mu = (-1,0,0,0)$. We can verify that this
transformation is done correctly in several ways. First, we can verify
that $u^\mu = (-1,0,0,0)$ after the transformation is done. This is
the case again to machine precision. 

More interesting tests of the frame transformation come from comparing
the combined redshift doppler shift factor $g$ found from
$-1/\hat{k}^t$ to that obtained from transforming a
generic momentum four-vector to the locally non-rotating frame
\citep{bardeen1972} for a generic four-velocity. The result is in
equation 17 in \citet{viergutz1993}, and a comparison to the method used
here is shown in the left panel of Fig. \ref{testrshiftang}. We find
good agreement at all points along the ray. In the right panel of
Fig. \ref{testrshiftang} we compare the angle between $k^\mu$ and
$b^\mu$ in the orthonormal fluid frame to the covariant method for
computing the same angle from \citet{broderick2004}: 

\begin{equation}
\cos^2 \theta_B = \frac{(b^\mu k_\mu)^2}{b^\nu b_\nu [k^\sigma k_\sigma + (k^\rho u_\rho)^2]}.
\end{equation}

\noindent The agreement is excellent, with significant deviations only appearing when $k^\mu b_\mu$ is very small.

We use the method of \citet{shcherbakovhuang2011} to project the local
polarisation basis in the fluid on to that of the parallel transported
polarisation basis of the observer. \citet{connorspiranstark1980} and
\citet{agolphd} used a similar method, but instead parallel
transported local vectors orthogonal and parallel to an accretion disc
in the equatorial plane and $k^\mu$ to the distant observer. For our purposes, we
want the orthogonal vector, which is given in Boyer-Lindquist
coordinates as \citep{agolphd},

\begin{eqnarray}
f^t_\perp &=& 0,\\
f^r_\perp &=& \frac{\sqrt{\Delta}k^{(\theta)} k^{(r)}}{r N_f},\\
f^\theta_\perp &=& \frac{1}{r N_f}\left[k^{(r)^2} + (1+v^2)
  k^{(\phi)^2}-2vk^{(\phi)} k^{(t)} + \right.\\&&\left.v k^{(\theta)^2} k^{(\phi)} / k^{(t)}\right],\\
f^\phi_\perp &=& \frac{r k^{(\theta)}}{\sqrt{A}
  N_f}\left[-(1+v^2)k^{(\phi)} + v k^{(t)} + \right.\\&&\left.v k^{(\phi)^2}/k^{(t)}\right],
\end{eqnarray}

\noindent where $k^{(\mu)}$ are the components of $k^\mu$ in the locally non-rotating
frame \citep{bardeen1972} and $N_f$ is a normalisation chosen so that
$f^\mu_\perp f_{\mu,\perp} = 1$. The polarisation angle $\psi$ is then
given in terms of $K_1$ and $K_2$ (equation \eqref{walkerpenrose}):

\begin{equation}
\tan \psi = \frac{-K_1 \beta-K_2 \gamma}{K_2 \beta - K_1\gamma}.
\end{equation}

We can directly compare this to the rotation angle $\chi$ from
equation \ref{chieq}, as long as we identify $b^\mu$, used in
\textsc{grtrans} as the polarisation reference vector, with
their disc normal vector, $f^\mu_\perp$ above. A comparison between
our angle $\chi$ and their $\psi$ is shown in Figure \ref{psichi} for
polarisation from electron scattering in a thin accretion disc. The
agreement is mostly good, although with deviations
$\lesssim 10\%$ near the event horizon.

These comparisons verify both sets of methods used for calculating
redshift/Doppler factors, and angles between the magnetic field and
wave vectors and between the polarisation basis in the fluid frame and
that of the observer, accounting for parallel transport along the
ray. The residual systematic errors $< 10\%$ in these
quantities, are comparable to the level of accuracy achieved in other
parts of the calculation (e.g. the integration or the transfer
coefficients).  

\subsection{Integration tests}
\label{sec:integration-tests}

\begin{figure}
\begin{center}
\includegraphics[scale=0.55]{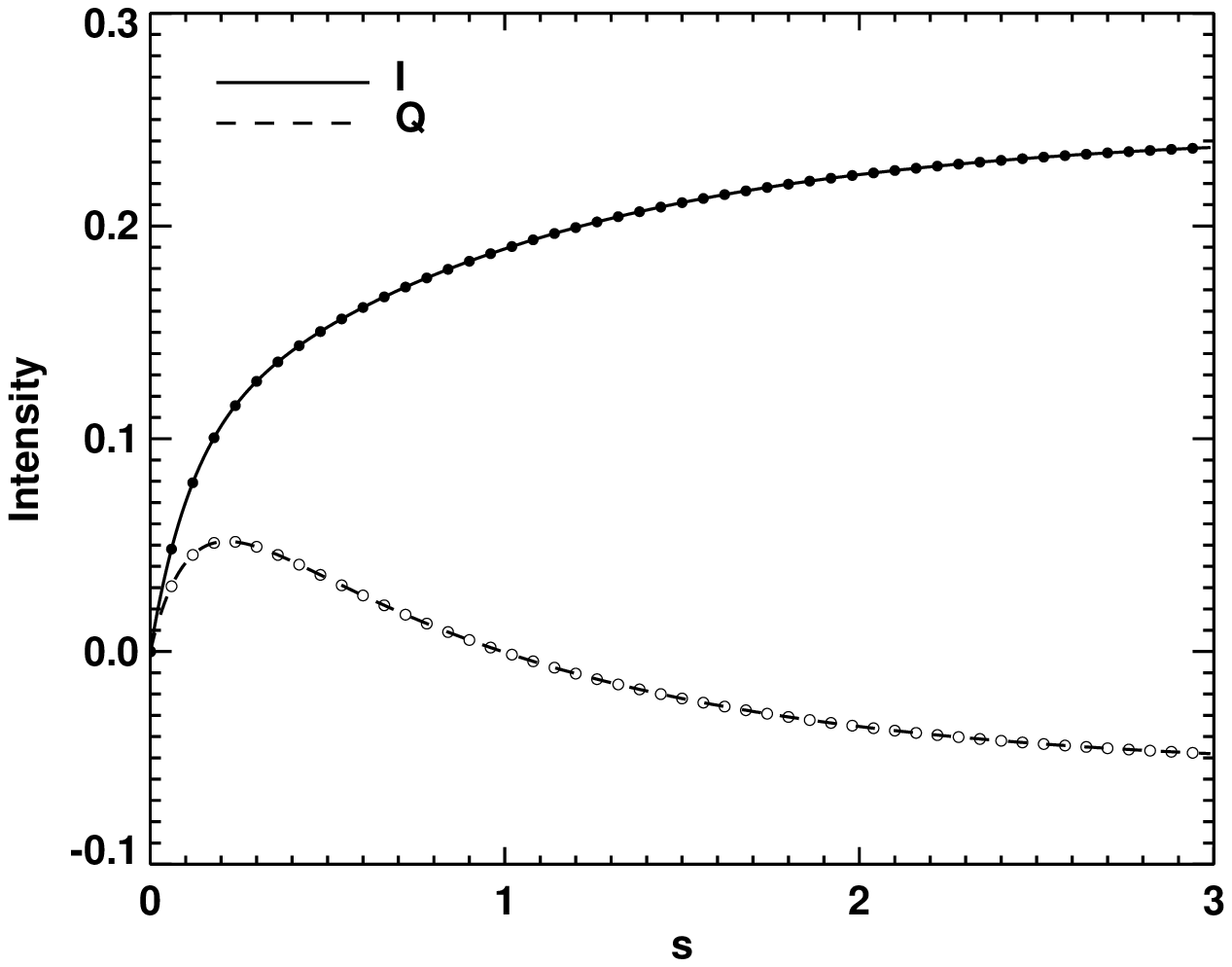}\\
\includegraphics[scale=0.55]{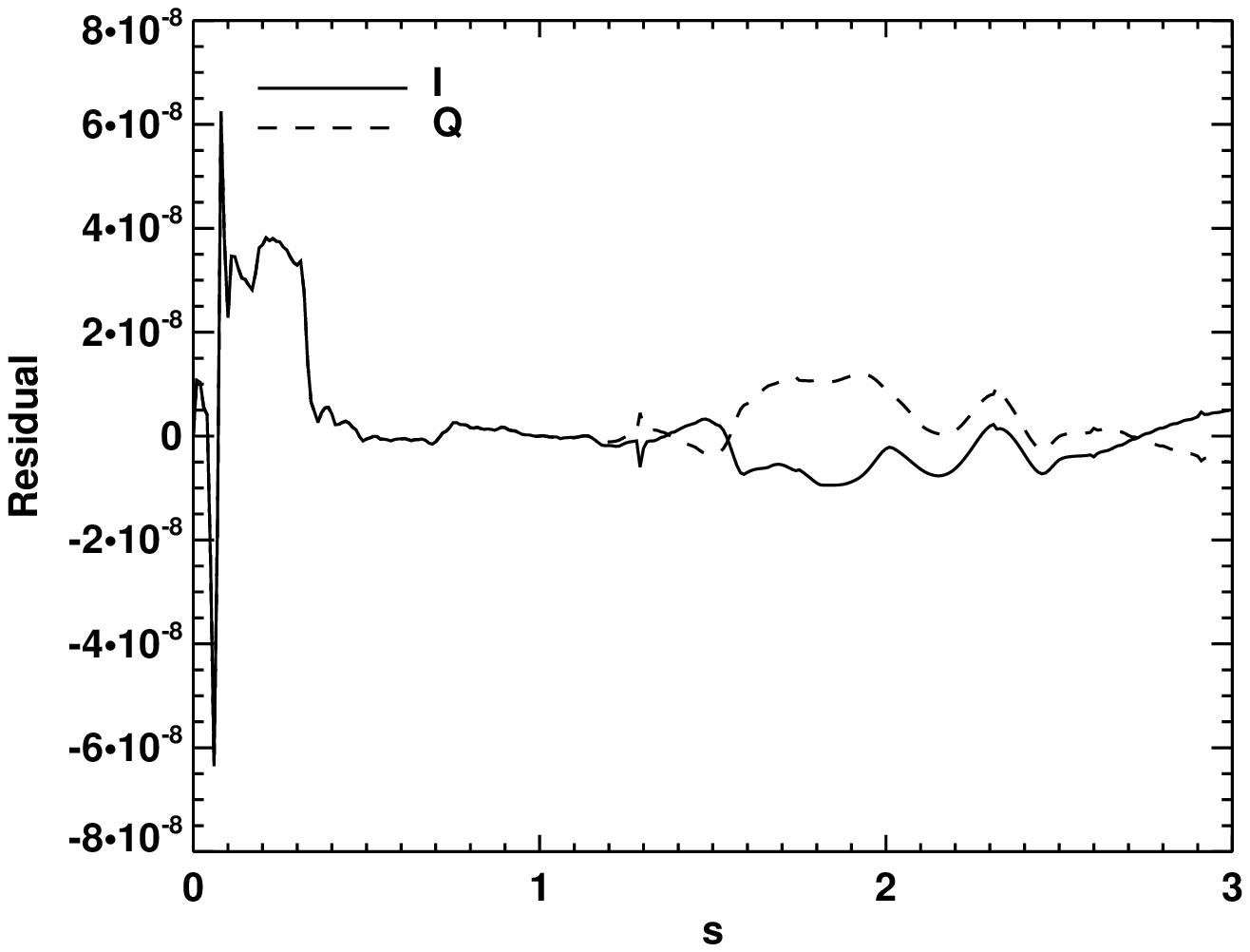}
\caption{\label{testabs} Analytic (lines) and numerical
  \textsc{grtrans} (dots)
  solutions to the polarised emission and absorption test problem (see
  \S \ref{sec:integration-tests}, Appendix \ref{sec:analyt-solut-polar}) for
  the Stokes parameters $I$ and $Q$. Single precision accuracy is
  maintained over the entire ray.}
\end{center}
\end{figure}

We test the accuracy and precision of the different methods for
integrating the polarised radiative transfer equations
\ref{sec:integration} through comparison to idealised, analytic
solutions with constant coefficients along a ray. We consider two test
problems, one for each limiting regime of the equations. The first
problem uses only emission and absorption in Stokes I,Q. The analytic
solution is given in equation \eqref{eq:2}, and a comparison of the analytic
solution and that calculated using the LSODA integration method is
shown in Figure \ref{testabs}. The agreement is excellent to within
single precision. 

\begin{figure}
\begin{center}
\includegraphics[scale=0.55]{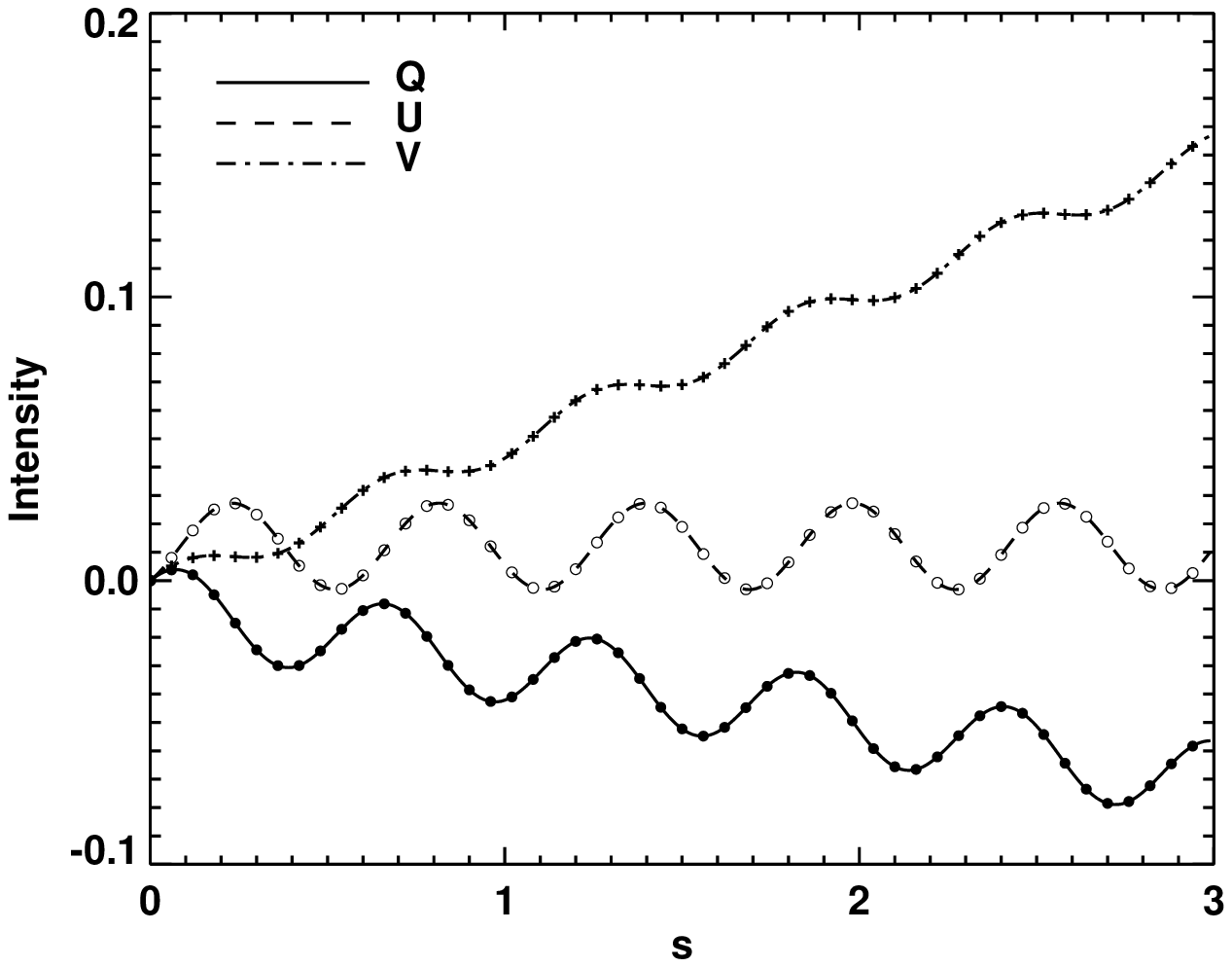}\\
\includegraphics[scale=0.55]{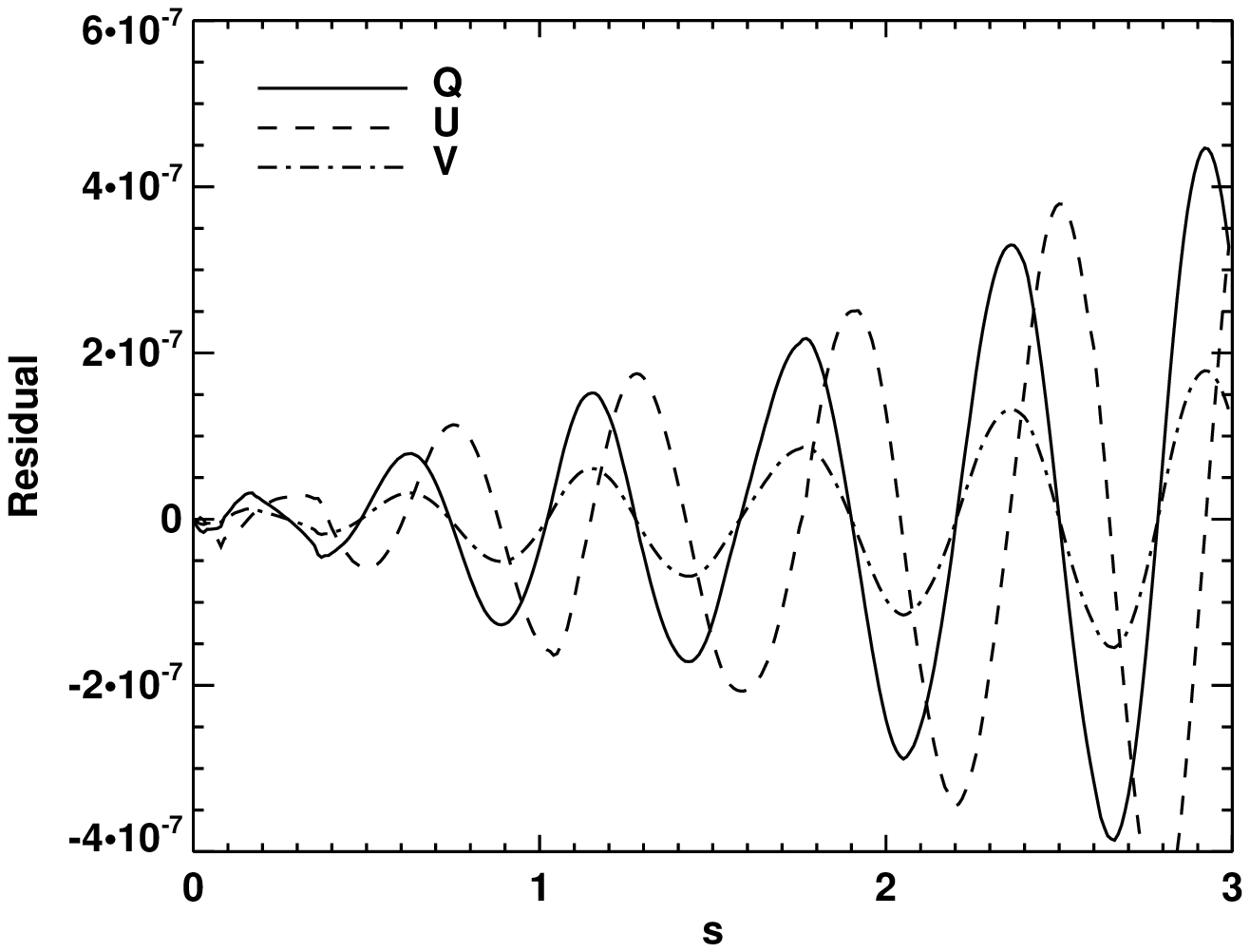}
\caption{\label{testfar} Analytic (lines) and numerical
  \textsc{grtrans} (dots)
  solutions to the intrinsic Faraday rotation and conversion test
  problem (see \S \ref{sec:integration-tests}, Appendix 
  \ref{sec:analyt-solut-polar}) for the Stokes parameters $Q$, $U$,
  and $V$. The residuals in this case grow along the ray. However, the absolute error remains small
unless a very large number of oscillations are present.}
\end{center}
\end{figure}

The second problem is the intensity in Stokes $Q$, $U$, $V$ for pure
Faraday rotation and conversion ($\rho_V$ and $\rho_Q$) with emission
in $Q$ and $V$. The analytic solution is purely oscillatory, and is
given in equation \eqref{eq:1}. Again the agreement between analytic and
numerical solutions is excellent (Fig. \ref{testfar}). In
this case the residuals grow with each oscillation. Still, the
absolute errors are so small that the error will be negligible unless
the Faraday optical depth is enormous, in which case code convergence
and run time will also become poor. This is not a limit of interest
here, but the issue and some possible solutions are discussed in 
\citet{shcherbakovetal2012}. 

\subsection{Test problems}
\label{sec:test-problems}

\begin{figure}
\begin{center}
\includegraphics[scale=0.55]{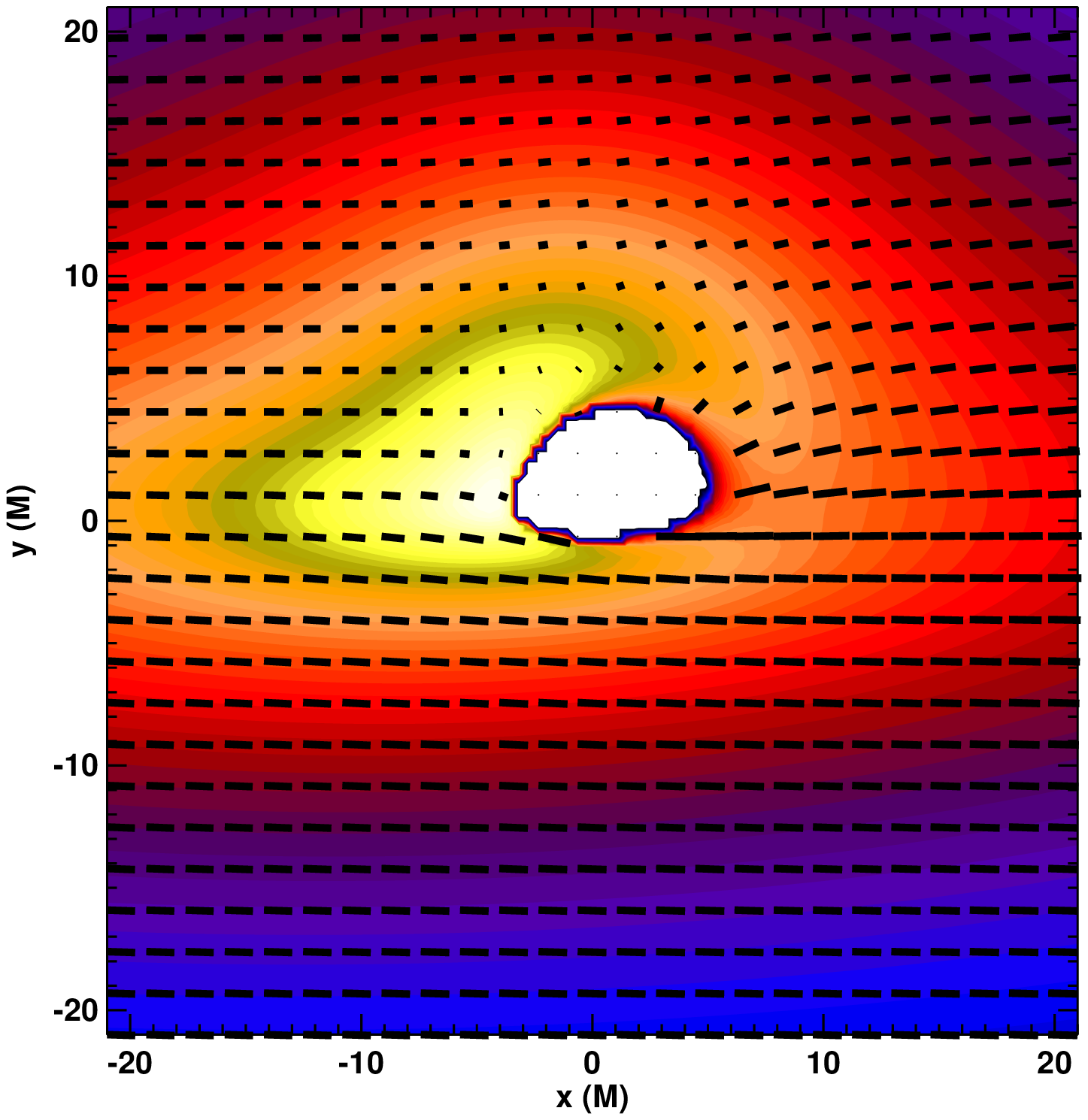}
\caption{\label{thindisc}Intensity and linear polarisation
  map of thermal emission from a thin accretion disc, ignoring the
  effect of returning radiation \citep{agolkrolik2000}. The intrinsic
  polarisation is assumed to follow the solution for scattering in a
  semi-finite atmosphere \citep{sobolev1963,chandrasekhar1950}. The black hole
  mass is $10 M_\odot$, the accretion rate is $0.1 \dot{M}_{\rm edd}$,
  and the inclination angle is $\theta_0 = 75^\circ$. The image is log-scaled
  with a color scale chosen to match Figure 1 of
  \citet{schnittmankrolik2009}. Both the total intensity and
  polarisation results are in excellent agreement with theirs.}
\end{center}
\end{figure}

Finally we show examples of full test problems based on calculations
in the literature. The first example is of the total intensity and
linear polarisation of a relativistic, thin accretion disc
\citep{page1974} in the equatorial plane. The emission is assumed to
be optically thick so that the emergent intensity from each point on
the disc is a blackbody at the local photospheric temperature. The
emergent polarisation is from electron scattering from a semi-infinite
slab \citep{sobolev1963,chandrasekhar1950}. Figure \ref{thindisc}
shows the resulting total intensity, on a log scale, and polarisation
vectors. The parameters are $M = 10 M_\odot$, $\dot{M} = 0.1
\dot{M}_{Edd}$, and the image is integrated over X-ray energies
$0.1-10$ keV. The results are in excellent agreement with Figure 1 of
\citet{schnittmankrolik2009}.

\begin{figure}
\begin{center}
\includegraphics[scale=0.55]{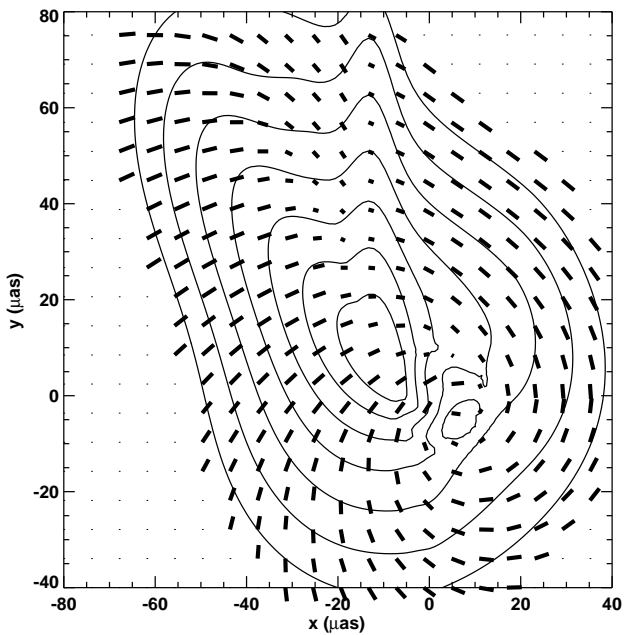}
\caption{\label{bl09} Total intensity (contours) and linear
  polarisation map from the semi-analytic jet model of
  \citet{broderickloeb2009}. The parameters are $a = 0.998$, $\xi =
  0.5$, $i = 25^\circ$, $\nu_0 = 345$ GHz. The results are
  mostly in good agreement with those of their M0 model (Figure
  7). The discrepancies arise in the counter-jet, which contains
  little of the total or polarised flux.}
\end{center}
\end{figure}

Next we calculate polarised synchrotron radiation from the semi-analytic jet model of \citet{broderickloeb2009}. The calculation of the jet structure is described there and in \S \ref{bl09jet}. The electrons in the jet are assumed to follow a power law distribution with a minimum Lorentz factor of $100$. Our transfer coefficients for this case are different than theirs, since we account for the cut off of the distribution function at low energies (see Appendix \ref{cha:polar-synchr-emiss}). The resulting total intensity and polarisation are shown in Figure \ref{bl09}, and are for the most part in good agreement with those of their M0 model in their Figure 7. The discrepancies are only in the polarisation structure of the counter-jet (bottom right of the image), which could be from differences in how the jet solution is reflected across the plane $z=0$. In any event that region of the image contains little total or polarised flux.

\begin{figure*}
\begin{center}
\includegraphics[scale=0.55]{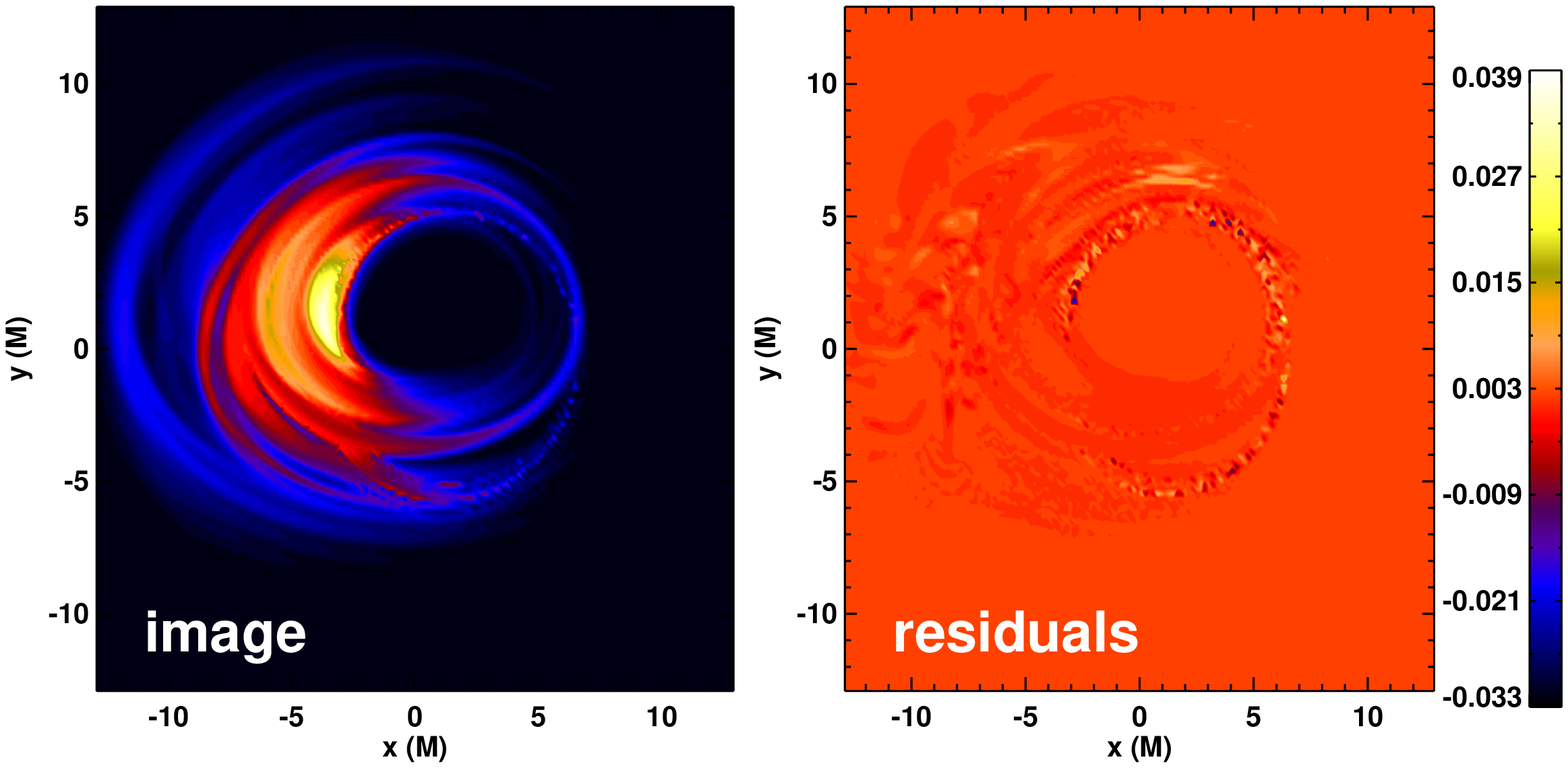}
\caption{\label{harm} Total intensity false color image of synchrotron
  radiation from a GRMHD simulation generated with the \textsc{HARM} code from
  the current version of \textsc{grtrans}, and the residuals between the
  current and previous versions of the code relative to the maximum
  overall image intensity. Only one image is shown, since they appear
  indistinguishable. The two code versions use 
  different methods for handling Doppler beaming, gravitational
  redshifts, and the integration of the radiative transfer
  equation. The agreement is excellent between the two cases: the
  maximum residuals in any pixel are $\sim 4\%$ of the
  maximum image intensity, and the total flux between the two cases
  agrees to $\simeq 0.3\%$.}
\end{center}
\end{figure*}

We can also compare the total intensity image from a relativistic MHD
simulation between the previous \citep{dexter2011,dexteretal2012proc}
and new versions of the \textsc{grtrans} code (Figure \ref{harm}). The
simulation used the public version of the \textsc{HARM} code
\citep{gammie2003,noble2006} with a black hole spin of $a =
0.9375$. The simulation results have been scaled to model the
submillimetre emission of Sagittarius A*, with a mean electron
temperature in the inner disc $\simeq 5\times10^{10}$ K and an
accretion rate chosen so that the flux at $\nu = 230$ GHz is roughly
$F_\nu \simeq 3$ Jy. The agreement between two independent versions of
the code is excellent (maximum pixel residuals $\simeq 4\%$ and total
flux residual $\simeq 0.3\%$). The previous version used the
alternative methods for finding $\theta_B$ and $g$ described in \S
\ref{sec:parall-transp-tests}, as well as a quadrature method for the
intensity. That code version also interpolated the fluid variables
rather than the emission and absorption coefficients. The residuals show that the systematic errors from
these different methods lead to only small difference in the resulting
total intensity image in a representative case.

\begin{figure*}
\begin{center}
\begin{tabular}{cc}
\includegraphics[scale=0.6]{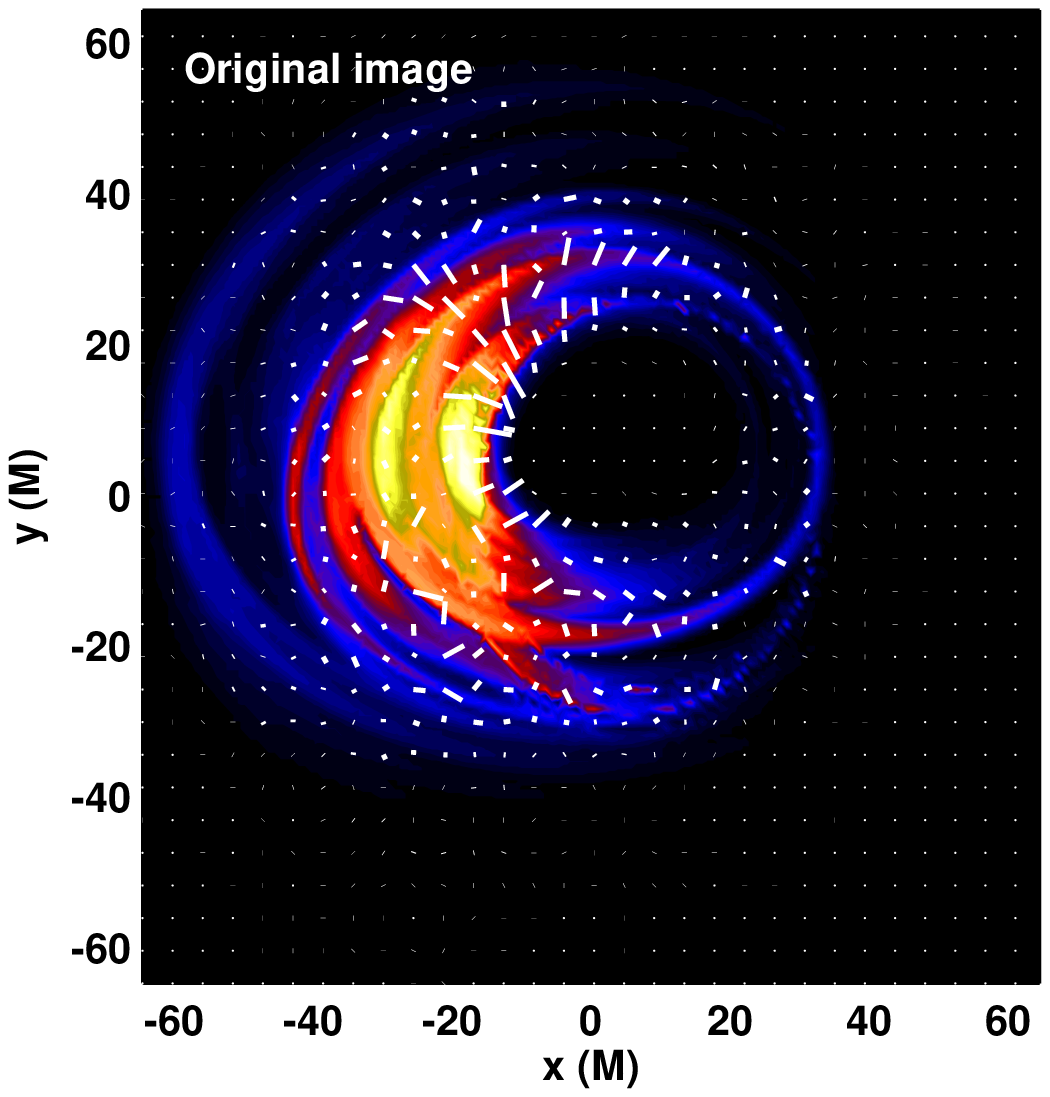}&
\includegraphics[scale=0.6]{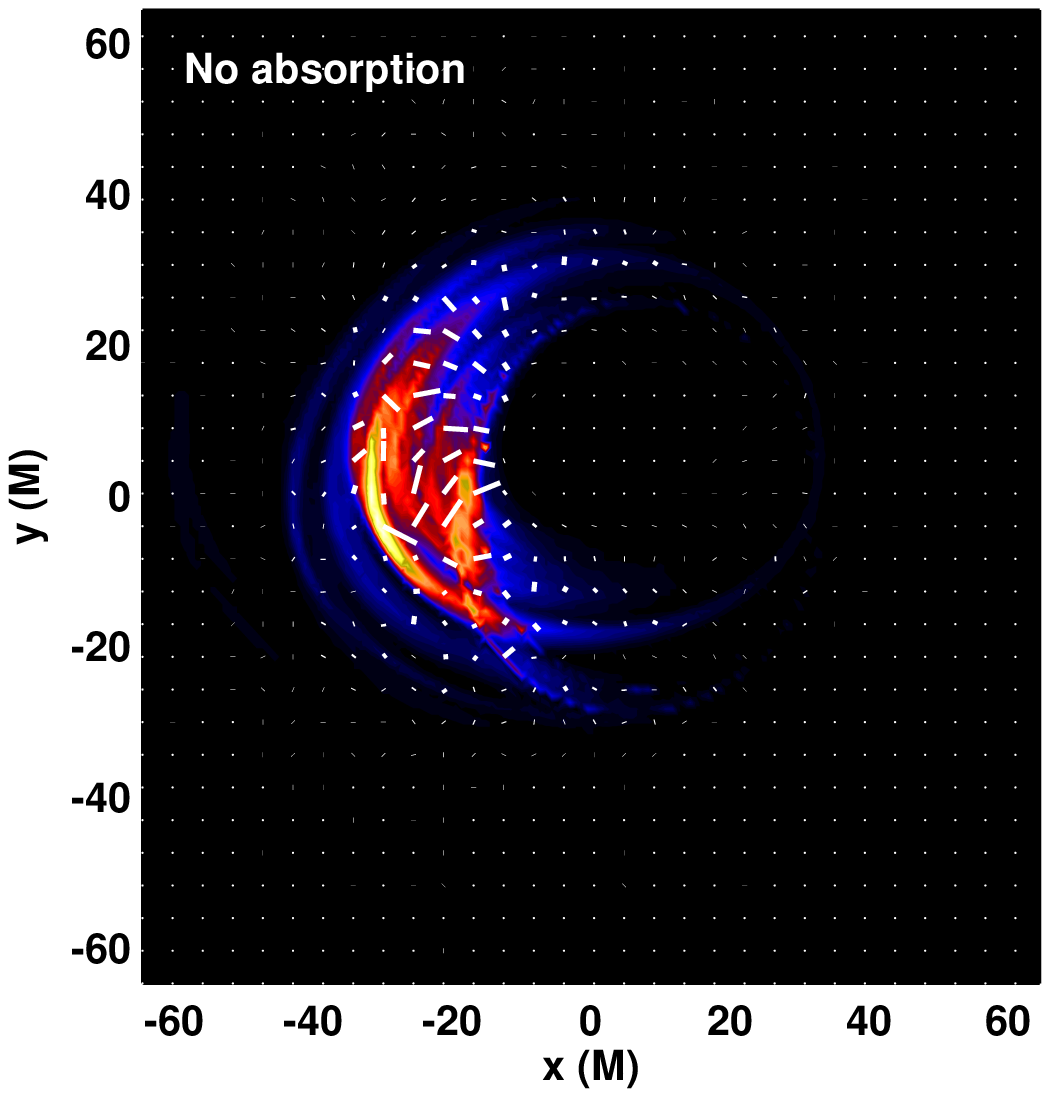}\\
\includegraphics[scale=0.6]{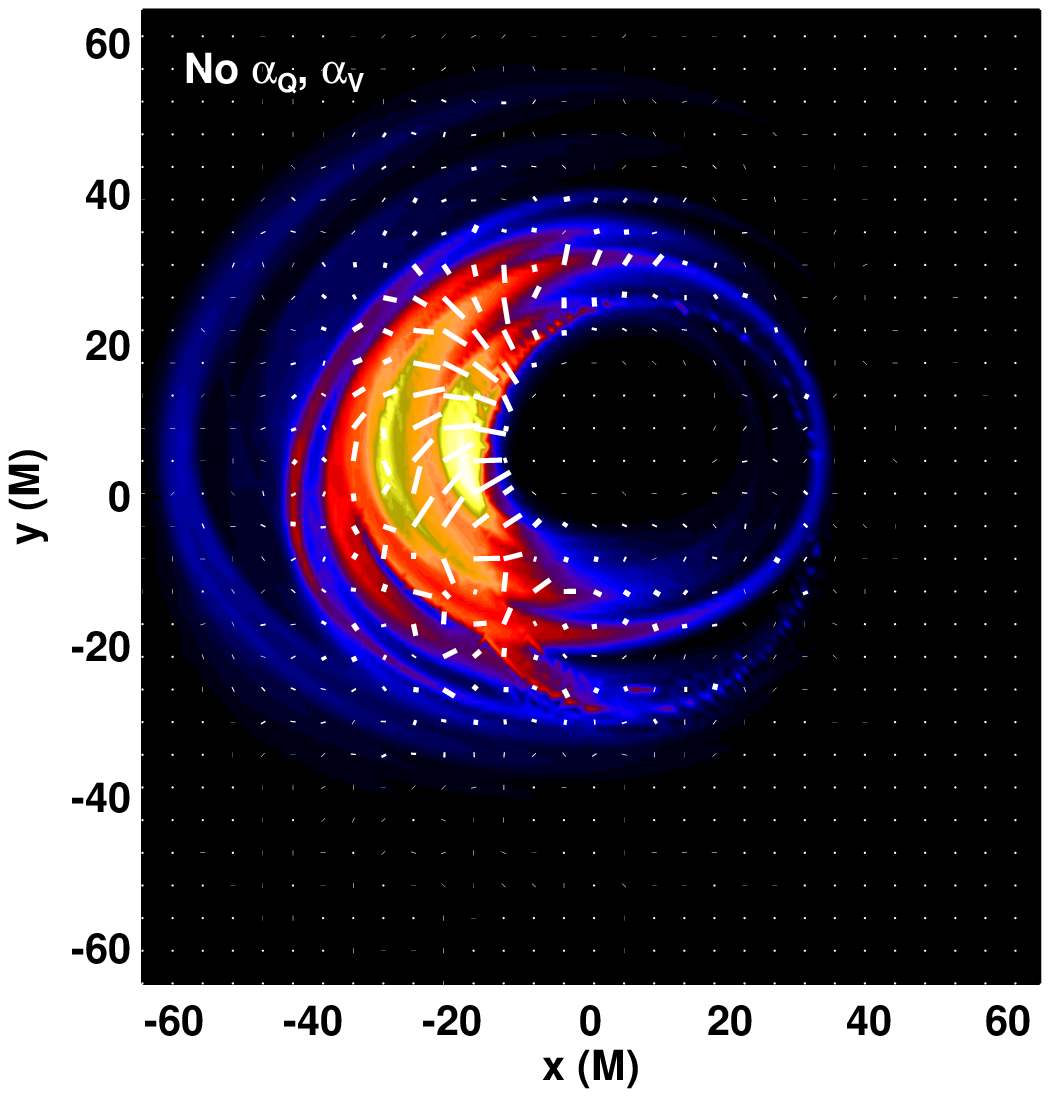}&
\includegraphics[scale=0.6]{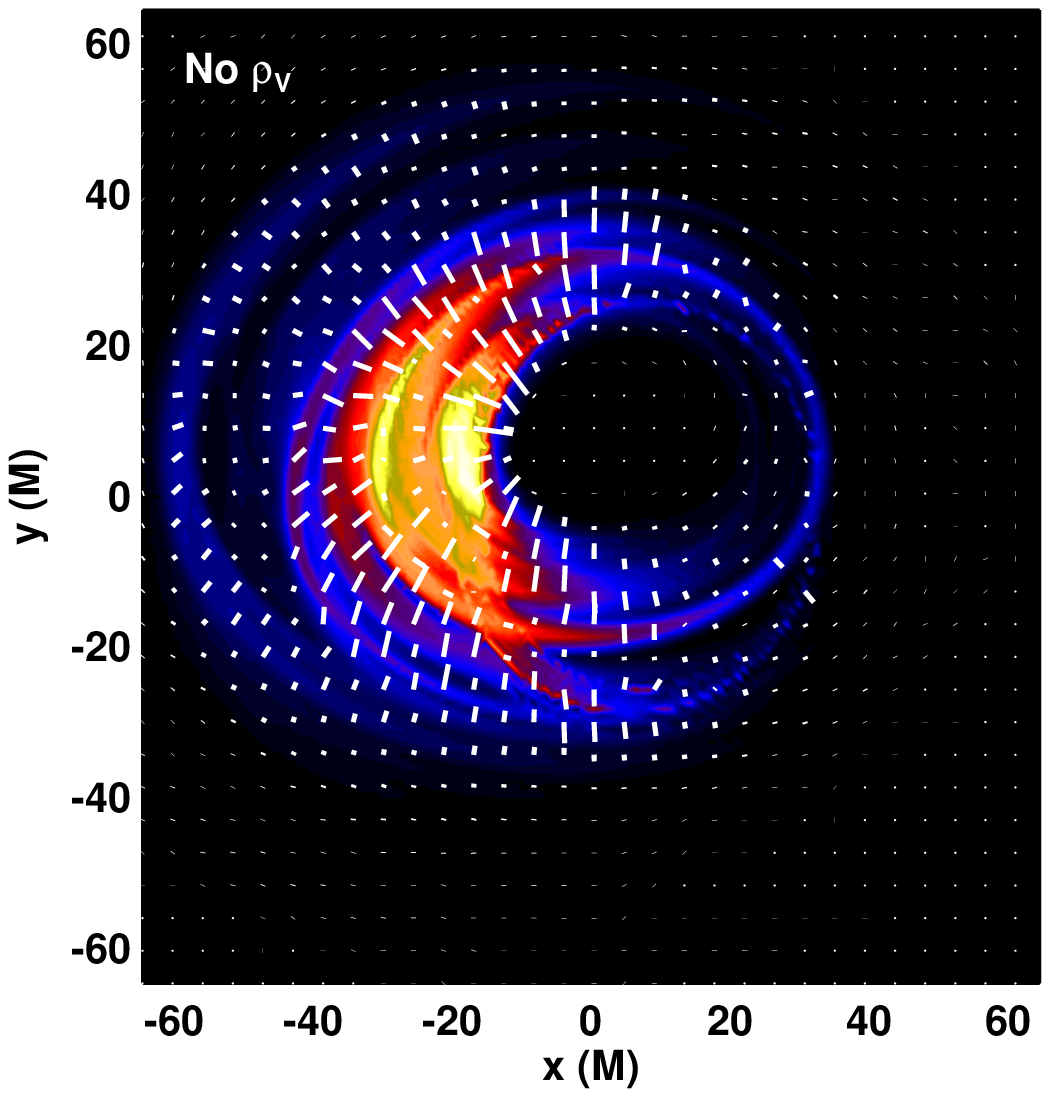}
\end{tabular}
\caption{\label{harmpol}Images and linear polarisation maps of Sgr A* corresponding to
  the \textsc{HARM} test problem. The top left image includes the full set of
  absorption and rotation coefficients. In each of the other panels,
  one or more of these coefficients are ignored to show their
  different effects on the total intensity and linear polarisation
  structure. Comparing the top
  left and bottom right images, for example, shows that intrinsic
  Faraday rotation is responsible for significantly de-polarising the
  resulting image. The
  polarised absorption components play an important role in 
  suppressing the polarisation in the brightest regions of the
  image. The top right panel shows that self-absorption plays an
  important role in both the total intensity and polarisation maps.}
\end{center}
\end{figure*}

As a final example, we show images and polarisation
maps from the \textsc{HARM} fluid model in Figure \ref{harmpol} with parameters
chosen to model the submm bump in Sgr A*
\citep[e.g.,][]{moscibrodzka2009,dexteretal2010}. The top left
panel includes all absorption and transfer effects. Including the
Faraday effects in particular leads to significant rotation of the
polarisation vectors and de-polarisation, in contrast to some previous
results finding coherent polarisation structures
\citep[e.g.,][]{bromley2001,broderickloeb2006} when Faraday effects
were ignored. The Faraday effects
arise within the emission region itself, even though the electrons are
mildly relativistic ($\theta_e \sim 10$).

We can understand this result in terms of known expressions for the
transfer coefficients (Appendix \ref{sec:farad-coeff-power}). The typical ratio $\nu /
\nu_c$ for these types of Sgr A* models in the submm is: 

\begin{equation}
\frac{\nu}{\nu_c} \simeq 40 \left(\frac{B}{30 \rm G}\right)^{-1}
\left(\frac{\theta_e}{10}\right)^{-2} \left(\frac{\nu}{230 \rm GHz}\right).
\end{equation}

\noindent At this value, for moderately relativistic temperatures the
Faraday coefficients can be much larger than the
total absorption coefficient (Figure
\ref{fig:rotcoef}). \citet{joneshardee1979} argued that because this
is only true when $\nu / \nu_c \gg 1$ where absorption is typically
negligible, Faraday rotation and conversion would be negligible
in thermal plasmas. However, the submm bump in Sgr A* is likely still
marginally self-absorbed
\citep[e.g.,][]{falcke1998,boweretal2015}. This is
certainly the case for these model images, where the image is
significantly modified in the top right panel when absorption is
neglected. In these models, the effective optical depth from Faraday
effects $\tau = \rho_{\rm Q, V} R \lesssim 100$ and therefore
significantly modifies the polarisation structure. Since Faraday effects
are sensitive to $\nu / \nu_c$ and $\theta_e$, the measured coherence
of the spatially resolved polarisation structure
\citep[e.g.,][]{johnsonetal2015} provides constraints on these quantities
and in turn on the properties of the emitting plasma. 

The polarised absorption coefficients also play a role in limiting the
polarisation fraction of the brightest pixels of the image (comparing
the top left and bottom left panels), but
including these components does not have a significant impact on the
total intensity image. Images of Sgr A* from previous calculations using only total
intensity radiative transfer are then unlikely to be subject
to systematic errors from neglecting these coefficients.

\section{Code structure and performance}
\label{sec:code-struct-perf}

In this section we describe the accuracy, convergence, performance,
and scaling of \textsc{grtrans}, and then provide a brief overview of
its organisation.

\subsection{Convergence}
\label{sec:convergence}

The accuracy of \textsc{grtrans} is very high for smooth solutions
(e.g., \S \ref{sec:integration-tests}), where the coefficients are tabulated over
much shorter sections of the ray than the intensity changes
appreciably. However, in problems of interest for ray tracing, the
emission and absorption coefficients generally change rapidly along
the ray, especially in the case of synchrotron radiation where they
are strong functions of the fluid state variables. In these cases the
rays will generally be sampled sparsely compared to the scale over
which the coefficients change. Then the
accuracy scales roughly linearly with the number of points along each
geodesic, and a sufficient number of points must be chosen to reach
the desired accuracy. 

\begin{figure*}
\begin{center}
\begin{tabular}{ll}
\includegraphics[scale=0.53]{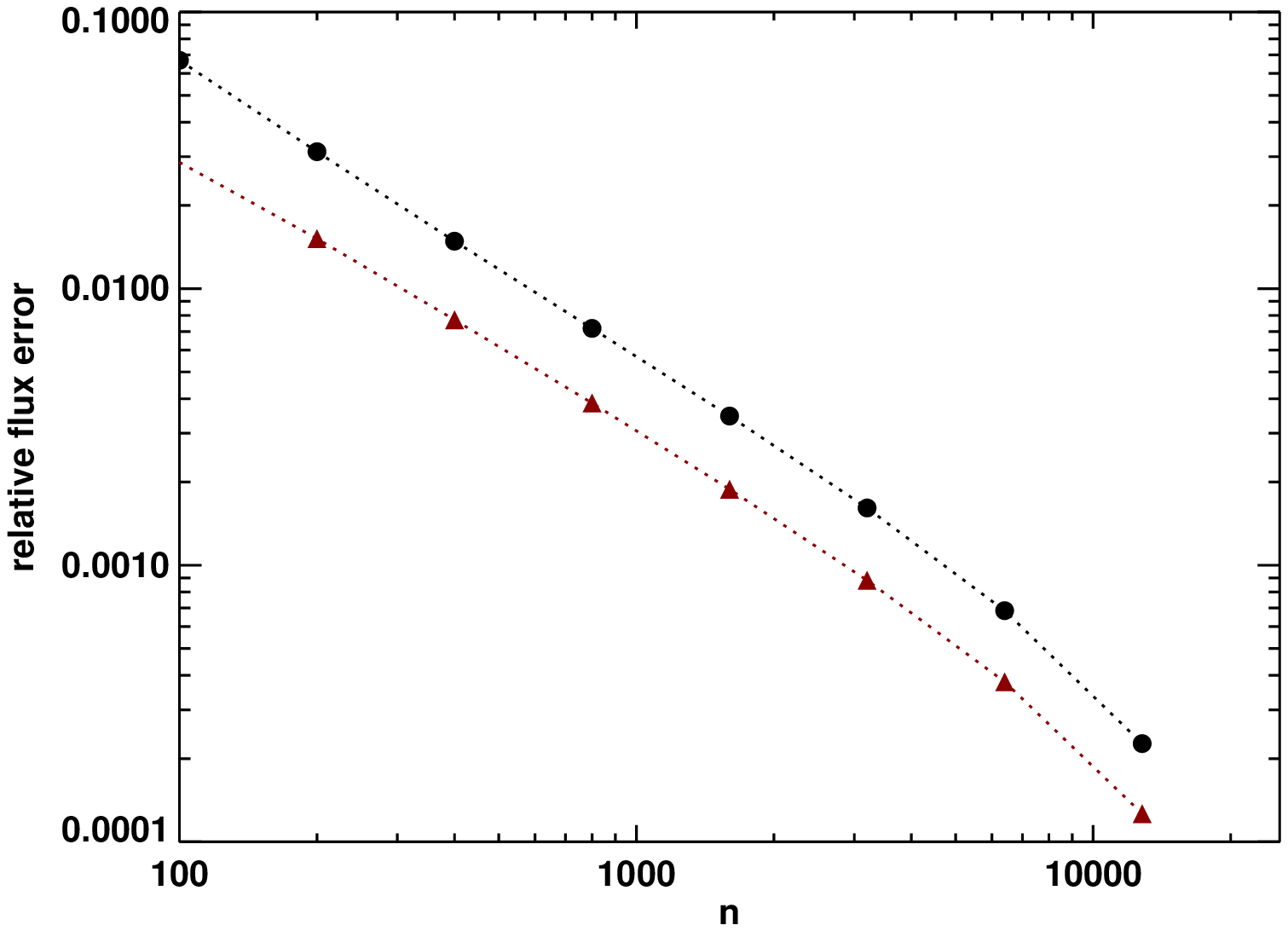}&
\includegraphics[scale=0.53]{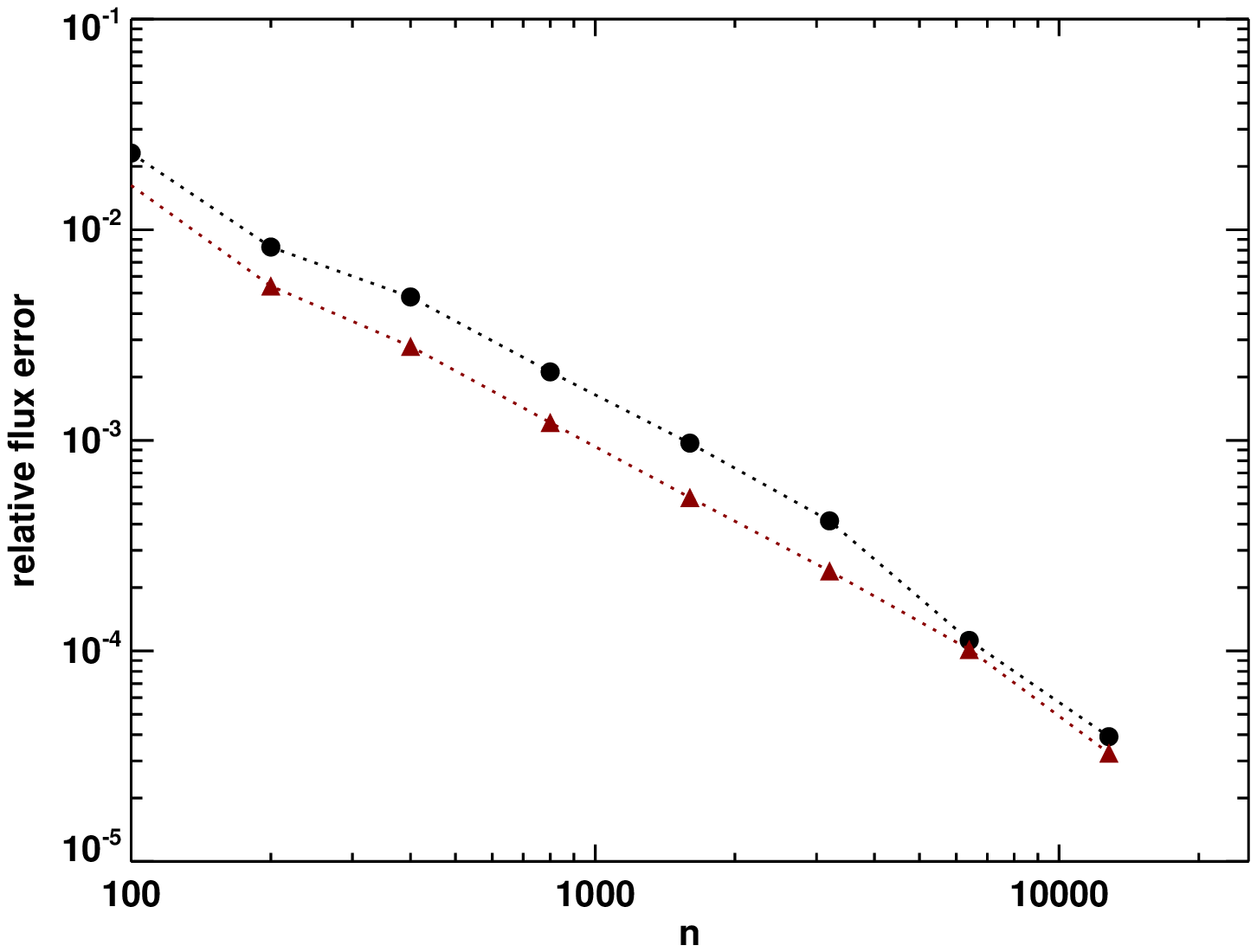}\\
\end{tabular}
\caption{\label{converge} Convergence of the total image flux as a function of the number of
  points tabulated along each geodesic for the M87 semi-analytic jet (left) and \textsc{HARM} Sgr A* (right) test
  problems with full polarisation using numerical (black circles) or
  quadrature (red triangles, equation \ref{omatrixmethod}) integration methods. The convergence is
  roughly linear with $n$, $\Delta F / F \sim 10^{-2} (n / 400)$, where the normalization of the error varies at order unity between different problems of interest, camera sizes, and integration methods.}
\end{center}
\end{figure*}

Figure~\ref{converge} shows the convergence of the total flux in the solutions to the \textsc{HARM} and semi-analytic jet 
problems as a function of $n$, compared to the solution with
$n=25600$. For typical problems of interest, the precision is better
than $\sim 1\% (n / 400)$. The precision is also usually better for
the numerical integration method than the formal solution method,
although there are cases where the reverse is true (bottom panel Figure \ref{converge}). In most
applications, $n >= 400$ should ensure that systematic errors
elsewhere in the code (e.g., in the approximations to the synchrotron
emissivities) would dominate the total error budget. There is no sign
of systematic disagreement between the two integration methods. With $n = 25600$,
their total fluxes agree to $0.01\%$, consistent with the linear
convergence of each method.

\subsection{Performance and scaling}
\label{sec:performance-scaling}

The calculation of the intensity at each camera pixel in ray tracing
are independent, and as such it is possible to speed up
calculations considerably on multi-core machines by assigning
different parts of the calculation to different cores. This is
achieved simply in \textsc{grtrans} by using different OpenMP threads
for different sets of camera pixels. Although there is
overhead associated with creating and destroying threads, the
efficiency is still high ($> 80\%$ in all problems and on all systems
studied), and with the added benefit that memory can be shared by all
threads, an important benefit for e.g. the post-processing of high
resolution 3D MHD simulations. Alternatively, threads could be used at
the level of different images, which might improve the efficiency, but
would then provide no speed up for calculating single images.

\begin{figure}
\begin{center}
\includegraphics[scale=0.55]{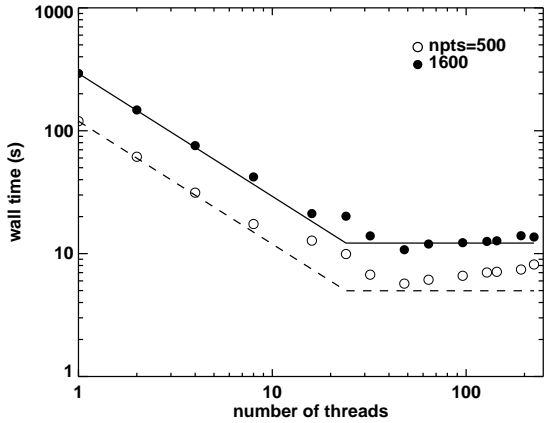}
\caption{\label{scalingabs} Strong scaling test of the
  \textsc{grtrans} code using the spherical accretion test problem. The wall time for
  single runs of the code with $n=500$ (open dots), $1600$ (solid dots) is plotted vs. the number of threads used on
a 24-core machine (2 12 core processors). The lines show 100\%
scaling for the machine based on the execution time for a single
thread. The peak efficiency of $\simeq 80-100\%$ for this problem is reached using 2
threads / core. The efficiency exceeds $100\%$ for 48 threads in the
$n=1600$ case, either due to run to run variability or improved
performance when hyperthreading is in use. The performance range found
here also applies to all other problems tested so far, and does not
seem to depend on the total number of cores or processors used.}
\end{center}
\end{figure}

Figure \ref{scalingabs} shows a strong scaling test for
\textsc{grtrans} using the spherical accretion example problem. The wall time taken by the parallel part of the code
is measured as a function of the number of OpenMP threads on a 24-core 
workstation. The
points are the measured times from single instances of running the
code, while the solid line is perfect scaling relative to the measured run time
using a single core. The efficiency for this problem peaks at 48
threads (2 threads / core or 1 thread / hyperthread), and is
$80-100\%$ for different values of $n$. These results are typical for
a wide range of test problems. 

\subsection{Code organisation}

The calculation of radiative transfer around a spinning black hole
consists of several independent pieces. In order to maintain flexibility, each of these
aspects of the calculation is implemented as a separate Fortran 90
module in \textsc{grtrans}. This Section describes the different
modules and how they are used together
to run \textsc{grtrans}. More detailed information about the code, 
explicit examples of its use, and guidelines for adding new fluid
and emission models are included in the code distribution.

\subsubsection{Kerr null geodesic calculation}

Rays in \textsc{grtrans} are assumed to
be null geodesics in a Kerr spacetime, and their trajectories in
Boyer-Lindquist coordinates are calculated using the semi-analytic
public code \textsc{geokerr} \citep{dexteragol2009}. In addition to
the existing public Fortran interfaces for \textsc{geokerr}, there is
now also a public Python interface to geokerr compiled using f2py.

The version of \textsc{geokerr} used by \textsc{grtrans} includes a
few minor bug fixes from the 
release version. The most important bug fix is that the option to use
$\mu = \cos \theta$ as an independent variable now works robustly even
when many turning points are present in a short segment of the orbit. A
bug associated with failures in the $\phi$ and $t$ coordinates in rare
cases where a ray is sampled extremely close to a turning point has
also been fixed. 

\subsubsection{Fluid models}
\label{sec:implement-fluids}

\textsc{grtrans} is designed to work with a range of models describing
the state variables of gas in the Kerr spacetime, from
non-relativistic semi-analytic solutions to the fluid equations
\citep[e.g.,][]{yuanquataert2003,broderick2009,broderickloeb2009} to
numerical solutions specified on a tabulated grid. These fluid models
are implemented separately, one per file, each of which contains a
common set of routines to initialise the model (including allocating
data), calculate fluid state variables at Boyer-Lindquist coordinate
positions in the Kerr metric, and delete the model (including
deallocating data). The code currently has several such models
implemented as are used in the example problems here. It is straightforward to
add new fluid models for use with the code using these existing models
as templates. 

Since the fluid models are implemented separately, they can be used
independently of \textsc{grtrans}. This is useful for testing that the
implementation is correct. Examples in the code are included also for
using f2py to build Python interfaces to such models, so that their
results can be accessed from Python. 

\subsubsection{Transfer coefficients}
\label{sec:implement-coefs}

In general, the calculation of the transfer coefficients is handled independently
of the fluid model. The emission models included at present are synchrotron
emission from thermal or power-law particle distributions and
optically thick color-corrected blackbody radiation, which can also
include linear polarisation induced from electron scattering in a
semi-infinite atmosphere.

As with fluid models, the user can include new emission models by
using the existing ones as templates. It is also straightforward to
combine various emissivities by writing a new one which then calls
combinations of those already in use. Examples of this included in the
code are the HYBRID and MAXJUTT emissivities, which are combinations of
synchrotron emission from thermal+PL and multiple thermal with
different temperatures (see Mao et al. 2015 for details).

The synchrotron emissivities can be compiled with f2py and used
directly from Python. 

\subsubsection{Other modules}

Many routines associated with the Kerr metric, including the
implementation of the method for parallel
transport of vectors along geodesics, are stored in their own
module. The integration methods for the radiative transfer equation
are as well, and also include an f2py interface for use in python.

\subsubsection{\textsc{grtrans} driver routine}

The main driver routine calculates the intensity at a specified number
of observed frequencies and values of other parameters (e.g. mass
accretion rate) for a given set of inputs.

The driver routine has global objects associated with the above
geodesic, fluid, emissivity, and radiative transfer modules. These are
used to store inputs and data. The objects are global so that they can
be accessed from the LSODA integration routines.

\subsubsection{Python interface}

A python class for \textsc{grtrans} includes
all of the code inputs and methods for reading the output. There are
two main interfaces to the code, either through the use of Fortran
input files (namelists) or through the Python wrapper to the code,
which compiles with f2py. Both interfaces can be used 
with Python, while the code can also be run from the
command line using input files.

\section{Discussion}

We have developed a new public code, \textsc{grtrans}, for polarised
ray tracing radiative transfer calculations in the Kerr metric,
designed with applications to modeling the emission from
low-luminosity black holes in mind. For this reason the code is currently focused on
synchrotron radiation (Appendix \ref{cha:polar-synchr-emiss}), and is
written to work with a wide range of underlying models for the
accreting or outflowing gas, from semi-analytic models (e.g., spherical accretion or
force-free jets) to relativistic MHD simulations (e.g., \textsc{HARM}). The
code is intended to be modular, so that it is straightforward to add
new fluid or emission models. It is
written in Fortran to make use of previous work on null geodesics and
other routines, but can be used efficiently from Python. We have
quantiatively compared results for independent methods for parallel
transport and integrating the polarised radiative 
transfer equations in an effort to verify the code, and presented full
examples of comparisons with published work. 

The code is written to do ray tracing in the Kerr metric, and as such
has two major limitations. First, many aspects of the code assume that
the background spacetime is the Kerr metric (e.g. the null geodesic
calculation in \textsc{geokerr} and the parallel transport
method). Generalising to other spacetimes is possible but would
require major changes to the code. The public \textsc{Gyoto} code
\citep{vincentetal2011} would probably be a better option for ray
tracing in a wide range of spacetimes, although at the moment it does
not include polarised radiative transfer. Second, ray tracing assumes
that the photon trajectories are known a priori, and so is impractical
for calculations where Compton scattering is important but where the
total Compton optical depth is small. In this case, one could
approximate the scattering locally, or do a first calculation to
estimate the effective emission/absorption from scattering. Still,
Monte Carlo methods such as those used in \textsc{grmonty}
\citep{dolenceetal2009} or
\textsc{Pandurata} \citep{schnittmankrolik2013} may be better suited
to such problems.

\section*{acknowledgements}
JD thanks S. Alwin Mao for significant contributions to the
development and testing of the code presented here, J. Davelaar and
M. Moscibrodzka for helpful feedback on the code and manuscript, and
C. Gammie for useful discussions. This work was
supported by a Sofja Kovalevskaja Award from the Alexander von
Humboldt Foundation of Germany.

\footnotesize{
\bibliographystyle{mn2e}
\bibliography{master}

\begin{thebibliography}{89}
\expandafter\ifx\csname natexlab\endcsname\relax\def\natexlab#1{#1}\fi

\bibitem[{{Abramowitz} \& {Stegun}(1970)}]{abramowitzstegun1970}
{Abramowitz} M., {Stegun} I.~A., 1970, {Handbook of mathematical functions :
  with formulas, graphs, and mathematical tables}

\bibitem[{{Agol}(1997)}]{agolphd}
{Agol} E., 1997, PhD thesis, University of California, Santa Barbara

\bibitem[{{Agol} \& {Krolik}(2000)}]{agolkrolik2000}
{Agol} E., {Krolik} J.~H., 2000, \apj, 528, 161

\bibitem[{{Akiyama} {et~al}\mbox{.}(2015){Akiyama}, {Lu}, {Fish}, {Doeleman},
  {Broderick}, {Dexter}, {Hada}, {Kino}, {Nagai}, {Honma}, {Johnson}, {Algaba},
  {Asada}, {Brinkerink}, {Blundell}, {Bower}, {Cappallo}, {Crew}, {Dexter},
  {Dzib}, {Freund}, {Friberg}, {Gurwell}, {Ho}, {Inoue}, {Krichbaum},
  {Loinard}, {MacMahon}, {Marrone}, {Moran}, {Nakamura}, {Nagar}, {Ortiz-Leon},
  {Plambeck}, {Pradel}, {Primiani}, {Rogers}, {Roy}, {SooHoo}, {Tavares},
  {Tilanus}, {Titus}, {Wagner}, {Weintroub}, {Yamaguchi}, {Young}, {Zensus}, \&
  {Ziurys}}]{akiyamaetal2015}
{Akiyama} K. {et~al.}, 2015, \apj, 807, 150

\bibitem[{{Balbus} \& {Hawley}(1991)}]{mri}
{Balbus} S.~A., {Hawley} J.~F., 1991, \apj, 376, 214

\bibitem[{{Bardeen}, {Press} \& {Teukolsky}(1972){Bardeen}, {Press}, \&
  {Teukolsky}}]{bardeen1972}
{Bardeen} J.~M., {Press} W.~H., {Teukolsky} S.~A., 1972, \apj, 178, 347

\bibitem[{{Beckwith}, {Hawley} \& {Krolik}(2008){Beckwith}, {Hawley}, \&
  {Krolik}}]{beckwith2008}
{Beckwith} K., {Hawley} J.~F., {Krolik} J.~H., 2008, \mnras, 390, 21

\bibitem[{{Blumenthal} \& {Gould}(1970)}]{blumenthalgould1970}
{Blumenthal} G.~R., {Gould} R.~J., 1970, Reviews of Modern Physics, 42, 237

\bibitem[{{Bower} {et~al}\mbox{.}(2015){Bower}, {Markoff}, {Dexter}, {Gurwell},
  {Moran}, {Brunthaler}, {Falcke}, {Fragile}, {Maitra}, {Marrone}, {Peck},
  {Rushton}, \& {Wright}}]{boweretal2015}
{Bower} G.~C. {et~al.}, 2015, \apj, 802, 69

\bibitem[{{Broderick} \& {Blandford}(2004)}]{broderickblandford2004}
{Broderick} A., {Blandford} R., 2004, \mnras, 349, 994

\bibitem[{{Broderick}(2004)}]{broderick2004}
{Broderick} A.~E., 2004, PhD thesis, California Institute of Technology,
  California, USA

\bibitem[{{Broderick} {et~al}\mbox{.}(2009){Broderick}, {Fish}, {Doeleman}, \&
  {Loeb}}]{broderick2009}
{Broderick} A.~E., {Fish} V.~L., {Doeleman} S.~S., {Loeb} A., 2009, \apj, 697,
  45

\bibitem[{{Broderick} \& {Loeb}(2005)}]{broderickloeb2005}
{Broderick} A.~E., {Loeb} A., 2005, \mnras, 363, 353

\bibitem[{{Broderick} \& {Loeb}(2006)}]{broderickloeb2006}
---, 2006, \apjl, 636, L109

\bibitem[{{Broderick} \& {Loeb}(2009)}]{broderickloeb2009}
---, 2009, \apj, 697, 1164

\bibitem[{{Bromley}, {Melia} \& {Liu}(2001){Bromley}, {Melia}, \&
  {Liu}}]{bromley2001}
{Bromley} B.~C., {Melia} F., {Liu} S., 2001, \apjl, 555, L83

\bibitem[{{Chan} {et~al}\mbox{.}(2015){Chan}, {Psaltis}, {{\"O}zel}, {Narayan},
  \& {Sa{\c d}owski}}]{chanetal2015}
{Chan} C.-K., {Psaltis} D., {{\"O}zel} F., {Narayan} R., {Sa{\c d}owski} A.,
  2015, \apj, 799, 1

\bibitem[{{Chandrasekhar}(1950)}]{chandrasekhar1950}
{Chandrasekhar} S., 1950, {Radiative transfer.} Oxford, Clarendon Press

\bibitem[{{Chandrasekhar}(1983)}]{chandrasekhar83}
---, 1983, {The mathematical theory of black holes}. Oxford/New York, Clarendon
  Press/Oxford University Press

\bibitem[{{Chen} {et~al}\mbox{.}(2015){Chen}, {Kantowski}, {Dai}, {Baron}, \&
  {Maddumage}}]{chenetal2015}
{Chen} B., {Kantowski} R., {Dai} X., {Baron} E., {Maddumage} P., 2015, \apjs,
  218, 4

\bibitem[{{Connors} \& {Stark}(1977)}]{connorsstark1977}
{Connors} P.~A., {Stark} R.~F., 1977, \nat, 269, 128

\bibitem[{{Connors}, {Stark} \& {Piran}(1980){Connors}, {Stark}, \&
  {Piran}}]{connorspiranstark1980}
{Connors} P.~A., {Stark} R.~F., {Piran} T., 1980, \apj, 235, 224

\bibitem[{{Cunningham}(1975)}]{cunningham1975}
{Cunningham} C.~T., 1975, \apj, 202, 788

\bibitem[{{Dauser} {et~al}\mbox{.}(2010){Dauser}, {Wilms}, {Reynolds}, \&
  {Brenneman}}]{dauseretal2010}
{Dauser} T., {Wilms} J., {Reynolds} C.~S., {Brenneman} L.~W., 2010, \mnras,
  409, 1534

\bibitem[{{Davis} \& {Hubeny}(2006)}]{davisetal2006}
{Davis} S.~W., {Hubeny} I., 2006, \apjs, 164, 530

\bibitem[{{De Villiers} \& {Hawley}(2003)}]{devilliers2003}
{De Villiers} J.-P., {Hawley} J.~F., 2003, \apj, 589, 458

\bibitem[{{Dexter}(2011)}]{dexter2011}
{Dexter} J., 2011, PhD thesis, University of Washington

\bibitem[{{Dexter} \& {Agol}(2009)}]{dexteragol2009}
{Dexter} J., {Agol} E., 2009, \apj, 696, 1616

\bibitem[{{Dexter} \& {Agol}(2011)}]{dexteragol2011}
---, 2011, \apjl, 727, L24

\bibitem[{{Dexter}, {Agol} \& {Fragile}(2009){Dexter}, {Agol}, \&
  {Fragile}}]{dexter2009}
{Dexter} J., {Agol} E., {Fragile} P.~C., 2009, \apjl, 703, L142

\bibitem[{{Dexter} {et~al}\mbox{.}(2010){Dexter}, {Agol}, {Fragile}, \&
  {McKinney}}]{dexteretal2010}
{Dexter} J., {Agol} E., {Fragile} P.~C., {McKinney} J.~C., 2010, \apj, 717,
  1092

\bibitem[{{Dexter} {et~al}\mbox{.}(2012){Dexter}, {Agol}, {Fragile}, \&
  {McKinney}}]{dexteretal2012proc}
---, 2012, Journal of Physics Conference Series, 372, 012023

\bibitem[{{Dexter} \& {Fragile}(2011)}]{dexterfragile2011}
{Dexter} J., {Fragile} P.~C., 2011, \apj, 730, 36

\bibitem[{{Dexter}, {McKinney} \& {Agol}(2012){Dexter}, {McKinney}, \&
  {Agol}}]{dexteretal2012}
{Dexter} J., {McKinney} J.~C., {Agol} E., 2012, \mnras, 421, 1517

\bibitem[{{Dexter} \& {Quataert}(2012)}]{dexterquataert2012}
{Dexter} J., {Quataert} E., 2012, \mnras, 426, L71

\bibitem[{{Doeleman} {et~al}\mbox{.}(2009){Doeleman}, {Agol}, {Backer},
  {Baganoff}, {Bower}, {Broderick}, {Fabian}, {Fish}, {Gammie}, {Ho}, {Honman},
  {Krichbaum}, {Loeb}, {Marrone}, {Reid}, {Rogers}, {Shapiro}, {Strittmatter},
  {Tilanus}, {Weintroub}, {Whitney}, {Wright}, \& {Ziurys}}]{doelemanetal2009}
{Doeleman} S. {et~al.}, 2009, in ArXiv Astrophysics e-prints, Vol. 2010,
  astro2010: The Astronomy and Astrophysics Decadal Survey, p.~68

\bibitem[{{Doeleman} {et~al}\mbox{.}(2012){Doeleman}, {Fish}, {Schenck},
  {Beaudoin}, {Blundell}, {Bower}, {Broderick}, {Chamberlin}, {Freund},
  {Friberg}, {Gurwell}, {Ho}, {Honma}, {Inoue}, {Krichbaum}, {Lamb}, {Loeb},
  {Lonsdale}, {Marrone}, {Moran}, {Oyama}, {Plambeck}, {Primiani}, {Rogers},
  {Smythe}, {SooHoo}, {Strittmatter}, {Tilanus}, {Titus}, {Weintroub},
  {Wright}, {Young}, \& {Ziurys}}]{doelemanetal2012}
{Doeleman} S.~S. {et~al.}, 2012, Science, 338, 355

\bibitem[{{Dolence} {et~al}\mbox{.}(2009){Dolence}, {Gammie},
  {Mo{\'s}cibrodzka}, \& {Leung}}]{dolenceetal2009}
{Dolence} J.~C., {Gammie} C.~F., {Mo{\'s}cibrodzka} M., {Leung} P.~K., 2009,
  \apjs, 184, 387

\bibitem[{{Eisenhauer} {et~al}\mbox{.}(2008){Eisenhauer}, {Perrin}, {Brandner},
  {Straubmeier}, {Richichi}, {Gillessen}, {Berger}, {Hippler}, {Eckart},
  {Sch{\"o}ller}, {Rabien}, {Cassaing}, {Lenzen}, {Thiel}, {Cl{\'e}net},
  {Ramos}, {Kellner}, {F{\'e}dou}, {Baumeister}, {Hofmann}, {Gendron}, {Boehm},
  {Bartko}, {Haubois}, {Klein}, {Dodds-Eden}, {Houairi}, {Hormuth},
  {Gr{\"a}ter}, {Jocou}, {Naranjo}, {Genzel}, {Kervella}, {Henning}, {Hamaus},
  {Lacour}, {Neumann}, {Haug}, {Malbet}, {Laun}, {Kolmeder}, {Paumard},
  {Rohloff}, {Pfuhl}, {Perraut}, {Ziegleder}, {Rouan}, \&
  {Rousset}}]{eisenhauer2008}
{Eisenhauer} F. {et~al.}, 2008, in Society of Photo-Optical Instrumentation
  Engineers (SPIE) Conference Series, Vol. 7013, Society of Photo-Optical
  Instrumentation Engineers (SPIE) Conference Series, p.~2

\bibitem[{{Falcke} {et~al}\mbox{.}(1998){Falcke}, {Goss}, {Matsuo}, {Teuben},
  {Zhao}, \& {Zylka}}]{falcke1998}
{Falcke} H., {Goss} W.~M., {Matsuo} H., {Teuben} P., {Zhao} J., {Zylka} R.,
  1998, \apj, 499, 731

\bibitem[{{Font}, {Ib{\'a}{\~n}ez} \& {Papadopoulos}(1999){Font},
  {Ib{\'a}{\~n}ez}, \& {Papadopoulos}}]{font1999}
{Font} J.~A., {Ib{\'a}{\~n}ez} J.~M., {Papadopoulos} P., 1999, \mnras, 305, 920

\bibitem[{{Gammie} \& {Leung}(2012)}]{gammieleung2012}
{Gammie} C.~F., {Leung} P.~K., 2012, \apj, 752, 123

\bibitem[{{Gammie}, {McKinney} \& {T{\'o}th}(2003){Gammie}, {McKinney}, \&
  {T{\'o}th}}]{gammie2003}
{Gammie} C.~F., {McKinney} J.~C., {T{\'o}th} G., 2003, \apj, 589, 444

\bibitem[{{Ginzburg} \& {Syrovatskii}(1965)}]{ginzburgsyrovatskii1965}
{Ginzburg} V.~L., {Syrovatskii} S.~I., 1965, \araa, 3, 297

\bibitem[{{Ginzburg} \& {Syrovatskii}(1969)}]{ginzburgsyrovatskii1969}
---, 1969, \araa, 7, 375

\bibitem[{{Gold} {et~al}\mbox{.}(2016){Gold}, {McKinney}, {Johnson}, \&
  {Doeleman}}]{goldmckinney2016}
{Gold} R., {McKinney} J.~C., {Johnson} M.~D., {Doeleman} S.~S., 2016, ArXiv
  e-prints

\bibitem[{{Hindmarsh}(1983)}]{odepack}
{Hindmarsh} A.~C., 1983, in Scientific Computing, {R. S. Stepleman et al.},
  ed., pp. 55--64

\bibitem[{{Huang} {et~al}\mbox{.}(2009){Huang}, {Liu}, {Shen}, {Yuan}, {Cai},
  {Li}, \& {Fryer}}]{huang2009}
{Huang} L., {Liu} S., {Shen} Z., {Yuan} Y., {Cai} M.~J., {Li} H., {Fryer}
  C.~L., 2009, \apj, 703, 557

\bibitem[{{Huang} \& {Shcherbakov}(2011)}]{huangshcherbakov2011}
{Huang} L., {Shcherbakov} R.~V., 2011, \mnras, 416, 2574

\bibitem[{{Johnson} {et~al}\mbox{.}(2015){Johnson}, {Fish}, {Doeleman},
  {Marrone}, {Plambeck}, {Wardle}, {Akiyama}, {Asada}, {Beaudoin}, {Blackburn},
  {Blundell}, {Bower}, {Brinkerink}, {Broderick}, {Cappallo}, {Chael}, {Crew},
  {Dexter}, {Dexter}, {Freund}, {Friberg}, {Gold}, {Gurwell}, {Ho}, {Honma},
  {Inoue}, {Kosowsky}, {Krichbaum}, {Lamb}, {Loeb}, {Lu}, {MacMahon},
  {McKinney}, {Moran}, {Narayan}, {Primiani}, {Psaltis}, {Rogers}, {Rosenfeld},
  {SooHoo}, {Tilanus}, {Titus}, {Vertatschitsch}, {Weintroub}, {Wright},
  {Young}, {Zensus}, \& {Ziurys}}]{johnsonetal2015}
{Johnson} M.~D. {et~al.}, 2015, Science, 350, 1242

\bibitem[{{Jones} \& {Hardee}(1979)}]{joneshardee1979}
{Jones} T.~W., {Hardee} P.~E., 1979, \apj, 228, 268

\bibitem[{{Jones} \& {Odell}(1977)}]{jonesodell1977}
{Jones} T.~W., {Odell} S.~L., 1977, \apj, 214, 522

\bibitem[{{Krolik}, {Hawley} \& {Hirose}(2005){Krolik}, {Hawley}, \&
  {Hirose}}]{kroliketal2005}
{Krolik} J.~H., {Hawley} J.~F., {Hirose} S., 2005, \apj, 622, 1008

\bibitem[{{Kulkarni} {et~al}\mbox{.}(2011){Kulkarni}, {Penna}, {Shcherbakov},
  {Steiner}, {Narayan}, {S{\"a} Dowski}, {Zhu}, {McClintock}, {Davis}, \&
  {McKinney}}]{kulkarnietal2011}
{Kulkarni} A.~K. {et~al.}, 2011, \mnras, 620

\bibitem[{{Landi Degl'Innocenti} \& {Landi
  Degl'Innocenti}(1985)}]{deglinnocenti1985}
{Landi Degl'Innocenti} E., {Landi Degl'Innocenti} M., 1985, \solphys, 97, 239

\bibitem[{{Legg} \& {Westfold}(1968)}]{leggwestfold1968}
{Legg} M.~P.~C., {Westfold} K.~C., 1968, \apj, 154, 499

\bibitem[{{Li} {et~al}\mbox{.}(2005){Li}, {Zimmerman}, {Narayan}, \&
  {McClintock}}]{lietal2005}
{Li} L.-X., {Zimmerman} E.~R., {Narayan} R., {McClintock} J.~E., 2005, \apjs,
  157, 335

\bibitem[{{Luminet}(1979)}]{luminet1979}
{Luminet} J.-P., 1979, \aap, 75, 228

\bibitem[{{Mahadevan}, {Narayan} \& {Yi}(1996){Mahadevan}, {Narayan}, \&
  {Yi}}]{maha}
{Mahadevan} R., {Narayan} R., {Yi} I., 1996, \apj, 465, 327

\bibitem[{{Melrose}(1971)}]{melrose1971}
{Melrose} D.~B., 1971, \apss, 12, 172

\bibitem[{{Melrose}(1980)}]{melrose1980}
---, 1980, {Plasma astrohysics. Nonthermal processes in diffuse magnetized
  plasmas - Vol.1: The emission, absorption and transfer of waves in plasmas;
  Vol.2: Astrophysical applications}. New York: Gordon and Breach, 1980

\bibitem[{{Melrose}(1997)}]{melrose1997}
---, 1997, Journal of Plasma Physics, 58, 735

\bibitem[{{Michel}(1972)}]{michel}
{Michel} F.~C., 1972, \apss, 15, 153

\bibitem[{{Moscibrodzka}, {Falcke} \& {Shiokawa}(2015){Moscibrodzka}, {Falcke},
  \& {Shiokawa}}]{moscibrodzkaetal2015}
{Moscibrodzka} M., {Falcke} H., {Shiokawa} H., 2015, ArXiv e-prints

\bibitem[{{Mo{\'s}cibrodzka} {et~al}\mbox{.}(2009){Mo{\'s}cibrodzka}, {Gammie},
  {Dolence}, {Shiokawa}, \& {Leung}}]{moscibrodzka2009}
{Mo{\'s}cibrodzka} M., {Gammie} C.~F., {Dolence} J.~C., {Shiokawa} H., {Leung}
  P.~K., 2009, \apj, 706, 497

\bibitem[{{Noble} {et~al}\mbox{.}(2006){Noble}, {Gammie}, {McKinney}, \& {Del
  Zanna}}]{noble2006}
{Noble} S.~C., {Gammie} C.~F., {McKinney} J.~C., {Del Zanna} L., 2006, \apj,
  641, 626

\bibitem[{{Noble} \& {Krolik}(2009)}]{noblekrolik2009}
{Noble} S.~C., {Krolik} J.~H., 2009, \apj, 703, 964

\bibitem[{{Noble} {et~al}\mbox{.}(2011){Noble}, {Krolik}, {Schnittman}, \&
  {Hawley}}]{nobleetal2011}
{Noble} S.~C., {Krolik} J.~H., {Schnittman} J.~D., {Hawley} J.~F., 2011, \apj,
  743, 115

\bibitem[{{Noble} {et~al}\mbox{.}(2007){Noble}, {Leung}, {Gammie}, \&
  {Book}}]{noble2007}
{Noble} S.~C., {Leung} P.~K., {Gammie} C.~F., {Book} L.~G., 2007, Class. and
  Quant. Gravity, 24, 259

\bibitem[{{Page} \& {Thorne}(1974)}]{page1974}
{Page} D.~N., {Thorne} K.~S., 1974, \apj, 191, 499

\bibitem[{{Rauch} \& {Blandford}(1994)}]{rauchblandford}
{Rauch} K.~P., {Blandford} R.~D., 1994, \apj, 421, 46

\bibitem[{{Rees}, {Durrant} \& {Murphy}(1989){Rees}, {Durrant}, \&
  {Murphy}}]{reesetal1989}
{Rees} D.~E., {Durrant} C.~J., {Murphy} G.~A., 1989, \apj, 339, 1093

\bibitem[{{Rybicki} \& {Lightman}(1979)}]{ryblight}
{Rybicki} G.~B., {Lightman} A.~P., 1979, {Radiative processes in astrophysics}.
  New York, Wiley-Interscience

\bibitem[{{Sazonov}(1969)}]{sazonov1969}
{Sazonov} V.~N., 1969, \sovast, 13, 396

\bibitem[{{Schnittman} \& {Krolik}(2009)}]{schnittmankrolik2009}
{Schnittman} J.~D., {Krolik} J.~H., 2009, \apj, 701, 1175

\bibitem[{{Schnittman} \& {Krolik}(2013)}]{schnittmankrolik2013}
---, 2013, \apj, 777, 11

\bibitem[{{Schnittman}, {Krolik} \& {Hawley}(2006){Schnittman}, {Krolik}, \&
  {Hawley}}]{schnittman2006}
{Schnittman} J.~D., {Krolik} J.~H., {Hawley} J.~F., 2006, \apj, 651, 1031

\bibitem[{{Shakura} \& {Sunyaev}(1973)}]{shaksun1973}
{Shakura} N.~I., {Sunyaev} R.~A., 1973, \aap, 24, 337

\bibitem[{{Shapiro}(1973{\natexlab{a}})}]{shap1}
{Shapiro} S.~L., 1973{\natexlab{a}}, \apj, 180, 531

\bibitem[{{Shapiro}(1973{\natexlab{b}})}]{shap2}
---, 1973{\natexlab{b}}, \apj, 185, 69

\bibitem[{{Shcherbakov}(2008)}]{shcherbakov2008}
{Shcherbakov} R.~V., 2008, \apj, 688, 695

\bibitem[{{Shcherbakov} \& {Huang}(2011)}]{shcherbakovhuang2011}
{Shcherbakov} R.~V., {Huang} L., 2011, \mnras, 410, 1052

\bibitem[{{Shcherbakov}, {Penna} \& {McKinney}(2012){Shcherbakov}, {Penna}, \&
  {McKinney}}]{shcherbakovetal2012}
{Shcherbakov} R.~V., {Penna} R.~F., {McKinney} J.~C., 2012, \apj, 755, 133

\bibitem[{{Sobolev}(1963)}]{sobolev1963}
{Sobolev} V.~V., 1963, {A treatise on radiative transfer.}

\bibitem[{{Viergutz}(1993)}]{viergutz1993}
{Viergutz} S.~U., 1993, \aap, 272, 355

\bibitem[{{Vincent} {et~al}\mbox{.}(2011){Vincent}, {Paumard}, {Gourgoulhon},
  \& {Perrin}}]{vincentetal2011}
{Vincent} F.~H., {Paumard} T., {Gourgoulhon} E., {Perrin} G., 2011, Classical
  and Quantum Gravity, 28, 225011

\bibitem[{{Walker} \& {Penrose}(1970)}]{walkerpenrose1970}
{Walker} M., {Penrose} R., 1970, Communications in Mathematical Physics, 18,
  265

\bibitem[{{Westfold}(1959)}]{westfold1959}
{Westfold} K.~C., 1959, \apj, 130, 241

\bibitem[{{Yuan}, {Quataert} \& {Narayan}(2003){Yuan}, {Quataert}, \&
  {Narayan}}]{yuanquataert2003}
{Yuan} F., {Quataert} E., {Narayan} R., 2003, \apj, 598, 301

\end{thebibliography}
}

\appendix
\onecolumn
\section{Polarised Synchrotron Emission and Absorption Coefficients for Thermal and Power Law Particle Distributions}
\label{cha:polar-synchr-emiss}

The subject of radiation from gyrating electrons in a magnetic field has been extensively studied, especially in the relativistic ``synchrotron'' limit where the electron energy $\gamma \gtrsim 1$  \citep{westfold1959,ginzburgsyrovatskii1965,ginzburgsyrovatskii1969,leggwestfold1968,sazonov1969,blumenthalgould1970,melrose1971,jonesodell1977,ryblight}. However, a consistent treatment of the derivation of the polarised emission and absorption coefficients for the two most commonly used particle distributions (thermal and power law) is still lacking. This appendix gives examples of deriving the various coefficients from integrating the single particle polarised synchrotron emissivity over distributions of particles and provides approximate formulae for their evaluation. The results are compared to emissivities found in the literature and in some cases to numerical integration.

The Stokes basis in the emitting frame has $\mathbf{B}=(0, 0, 1)$, $\mathbf{e^1}=(-\cos{\theta_B}, 0, \sin{\theta_B})$ and $\mathbf{e^2}=(0, 1, 0)$ where $\theta_B$ is the angle between $B$ and the wave-vector $k$ and $\mathbf{e^1}$, $\mathbf{e^2}$ are aligned with Stokes $Q$ and $U$ and the projection of $B$ onto the Stokes basis is entirely along $\mathbf{e^2}$. Then the vacuum emissivity can be written as a rank-2 tensor \citep[e.g.][]{melrose1980}:

\begin{equation}
\eta^{\alpha \beta} = \frac{\sqrt{3} e^2}{8\pi c} \nu_B \sin{\theta_B} H^{\alpha \beta} (\nu, \theta_B),
\end{equation}

\noindent where $e$ is the electron charge, $c$ is the speed of light, $\nu_B=\frac{eB}{2\pi m c}$, and  

\begin{eqnarray}
H^{11}&=&F\left(\frac{\nu}{\nu_c}\right)-G\left(\frac{\nu}{\nu_c}\right),\\
H^{22}&=&F\left(\frac{\nu}{\nu_c}\right)+G\left(\frac{\nu}{\nu_c}\right),\\
H^{12}&=&-H^{21}=\frac{4i\cot{\theta_B}}{3\gamma} H\left(\frac{\nu}{\nu_c}\right),
\end{eqnarray}

\noindent where $\nu$ is the emitted frequency, $\gamma$ is the
electron Lorentz factor, $\nu_c=3/2\nu_B \sin{\theta_B} \gamma^2$ and

\begin{eqnarray}
F(x)&=&x \int_x^\infty dy K_{5/3} (y),\\
G(x)&=&x K_{2/3} (x),\\
H(x)&=&\int_x^\infty dy K_{1/3} (y) + x K_{1/3} (x),
\end{eqnarray}

\noindent are the synchrotron functions for total, linearly and circularly polarised emission respectively and $K_\alpha (z)$ is the modified Bessel function.

To compute the emissivity from a distribution of electrons, these formulae are integrated over the particle distribution:

\begin{equation}\label{jalphabeta}
j^{\alpha \beta} = \int_0^\infty d\gamma N(\gamma) \eta^{\alpha \beta}.
\end{equation}

\noindent The Stokes emissivities are then given as
$j_I=j^{22}+j^{11}$, $j_Q=j^{22}-j^{11}$, $j_U=j_{12}+j_{12}$, and
$j_V=i(j_{12}-j_{21})$. For this Stokes basis, $j_U$ vanishes.

The two most commonly used particle distributions for astrophysical sources are the relativistic thermal (Maxwell) distribution, 

\begin{equation}\label{thermal}
N(\gamma)=\frac{n\gamma^2\beta \exp{(-\gamma/\theta_e)}}{\theta_e K_2(1/\theta_e)}
\end{equation}

\noindent where $n$ is the electron number density and $\theta_e=\frac{kT}{mc^2}$ is the dimensionless electron temperature; and the power law distribution,

 \begin{displaymath}
   N(\gamma) = \left\{
     \begin{array}{lr}\label{pl}
       n (p-1) (\gamma_1^{1-p}-\gamma_2^{1-p})^{-1} \gamma^{-p} & \gamma_1 < \gamma < \gamma_2\\
       0 & \mathrm{otherwise}
     \end{array}
   \right.
\end{displaymath} 

\noindent where $\gamma_{1,2}$ are the low- and high-energy cutoffs of the distribution.

We consider these two cases in turn and derive approximate formulae for their evaluation.

\subsection{Ultrarelativistic Thermal Distribution}

For the thermal distribution, substituting equation \eqref{thermal} into equation \eqref{jalphabeta} with $\beta \simeq 1$ and $\theta_e \gg 1$ gives,

\begin{equation}
j^{\alpha \beta} = \frac{\sqrt{3} n e^2 \nu_B \sin{\theta_B}}{8\pi\theta_e (2\theta_e^2)}\int_0^\infty d\gamma \gamma^2 \exp{(-\gamma/\theta_e)} H^{\alpha \beta} (\nu, \theta_B),
\end{equation}

\noindent where the approximate form of the modified bessel function
for small argument $K_2 (z) \rightarrow 2 z^2$ was used. First
substitute $z \equiv \gamma/\theta_e$ so that,

\begin{equation}
j^{\alpha \beta}=\frac{\sqrt{3} n e^2 \nu_B \sin{\theta_B} \theta_e^2}{8\pi c (2 \theta_e^2)}\int_0^\infty dz z^2 \exp{(-z)} H^{\alpha\beta} (\nu, \theta_B).
\end{equation}

\noindent Then substitute $\gamma$ for $z$ in the synchrotron functions and use the relations between $j^{\alpha\beta}$ and $j_{I,Q,V}$ to find:

\begin{eqnarray}
j_{I} (\nu, \theta_B) &=& \frac{n e^2 \nu}{2\sqrt{3} c \theta_e^2} I_I(x), \\
j_{Q} (\nu, \theta_B) &=& \frac{n e^2 \nu}{2\sqrt{3} c \theta_e^2} I_Q(x), \\
j_{V} (\nu, \theta_B) &=& \frac{2 n e^2 \nu \cot{\theta_B}}{3\sqrt{3} c \theta_e^3} I_V(x),
\end{eqnarray}

\noindent where $x \equiv \nu/\nu_c$ and here $\theta_e$ takes the
place of $\gamma$ in the definition of $\nu_c$, and the thermal synchrotron integrals are,

\begin{eqnarray}
I_I(x)&=&\frac{1}{x}\int_0^\infty dz z^2 \exp{(-z)} F\left(\frac{x}{z^2}\right), \\
I_Q(x)&=&\frac{1}{x}\int_0^\infty dz z^2 \exp{(-z)} G\left(\frac{x}{z^2}\right), \\
I_V(x)&=&\frac{1}{x}\int_0^\infty dz z \exp{(-z)} H\left(\frac{x}{z^2}\right).
\end{eqnarray}

\noindent where the function $I_I (x)$ corresponds to $I(x_M)$
from \citet{maha}. This result agrees with the formulae from previous work
\citep{sazonov1969,maha,huang2009}. The integrals can be approximated
analytically with high accuracy by matching the asymptotic behavior
for small and large arguments and fitting polynomials in the
transition region \citep{maha}. We find the following approximate forms,

\begin{eqnarray}
I_I(x)&=&2.5651(1+1.92x^{-1/3}+0.9977x^{-2/3})\exp{(-1.8899x^{1/3})},\\
I_Q(x)&=&2.5651(1+0.932x^{-1/3}+0.4998x^{-2/3})\exp{(-1.8899x^{1/3})},\\
I_V(x)&=&(1.8138x^{-1}+3.423x^{-2/3}+0.02955x^{-1/2}+2.0377x^{-1/3})\exp{(-1.8899x^{1/3})},
\end{eqnarray}

\noindent all agree with numerical integration within $\lesssim
1\%$ for all $x$. We further compare the results to numerical
integration of the full emissivities using the public
\textsc{symphony}\footnote{https://github.com/afd-illinois/symphony} 
code (Pandya et al. 2016). All fitting functions are accurate to within $\lesssim
20\%$ for parameters of interest ($\theta_e > 3$, $\nu / \nu_c >
1$), but our circular polarization emissivity has larger
deviations at low temperature ($\theta_e < 1$).

The absorption coefficients are computed from the emission
coefficients assuming local thermodynamic equilibrium so that
Kirchoff's Law, $j_\nu=\alpha_\nu B_\nu$, holds with $B_\nu$ the
blackbody function \citep[e.g.][]{ryblight}. 

\subsection{Power Law Distribution}

\begin{figure}
\begin{tabular}{ll}
\includegraphics[scale=0.65]{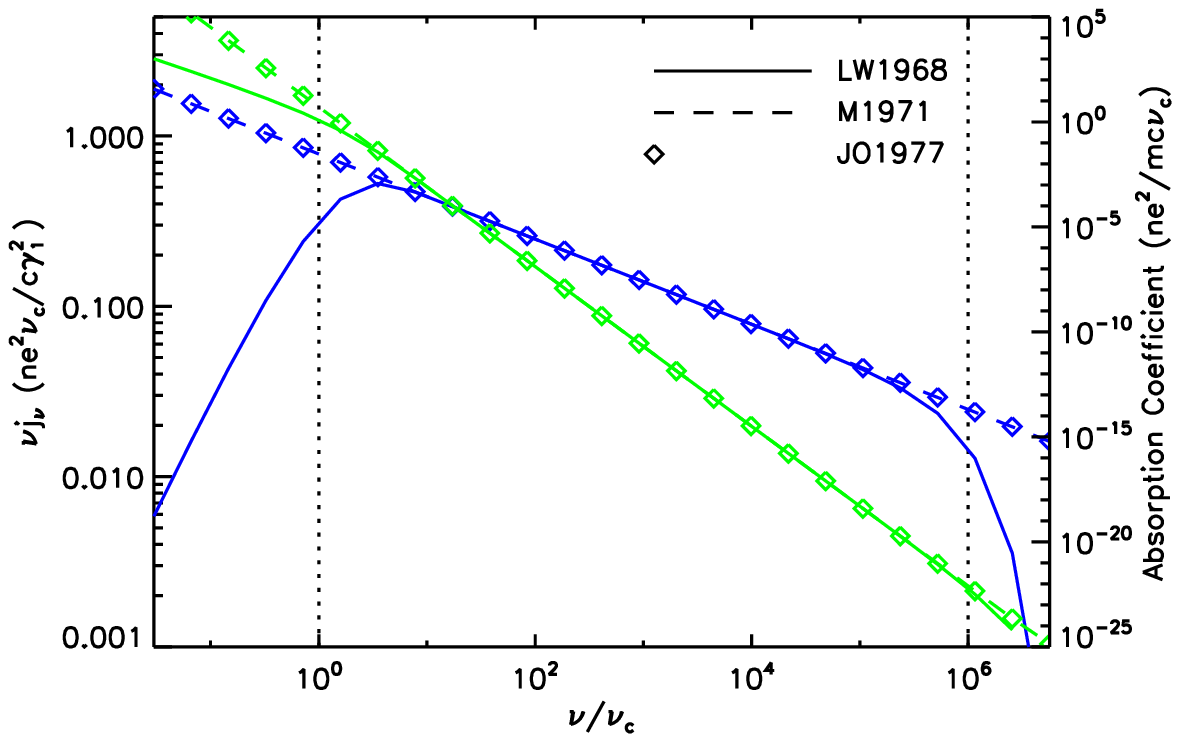}&
\includegraphics[scale=0.65]{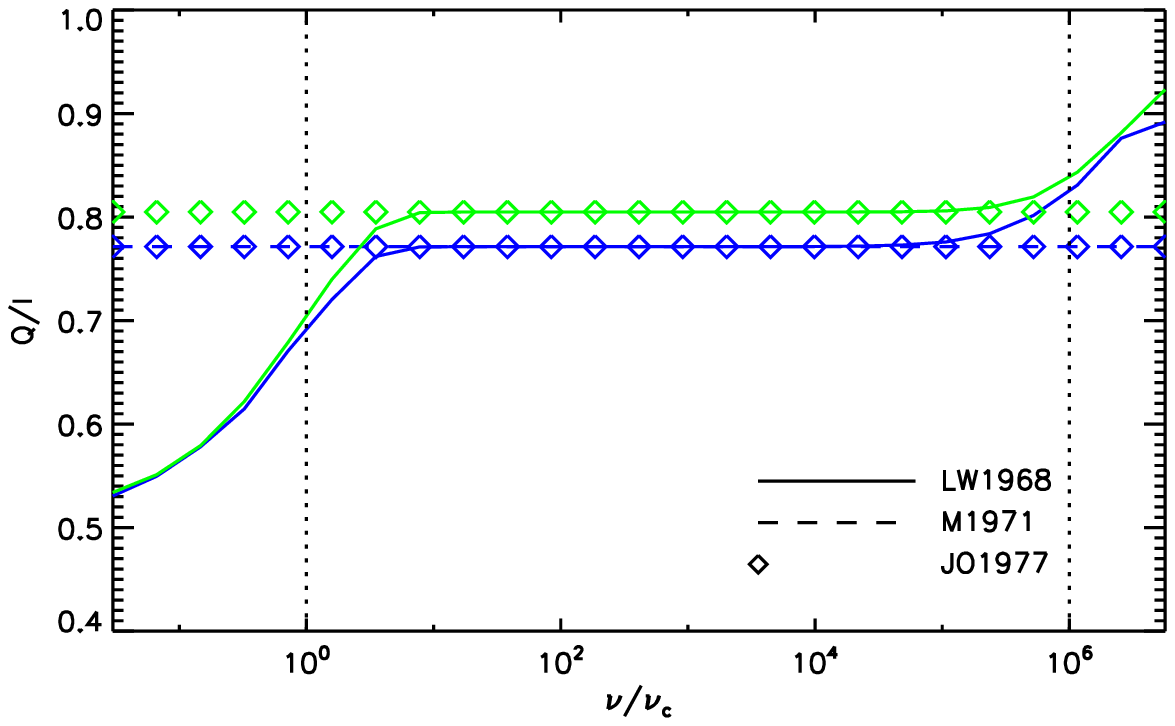}
\end{tabular}
\caption{\label{plcompare}\emph{Left:} total emission (blue) and absorption (green)
coefficients from this work and Legg \& Westfold (solid), Melrose (dashed) and
Jones \& Odell (diamonds). \emph{Right:} Linear polarisation fraction in emission (blue) and absorption (green)
coefficients from this work and Legg \& Westfold (solid), Melrose
(dashed) and Jones \& Odell (diamonds). Here, $\nu_c \equiv \nu_p
\gamma_1^2$ and the dotted lines show the locations of $\nu / \nu_c =
1$, $\gamma_2^2 / \gamma_1^2$, where the approximate forms of the
coefficients lose accuracy.}
\end{figure}

In this case, after plugging in the distribution we change the variable of integration to $x \equiv
\nu/\nu_c$:

\begin{equation}
j^{\alpha \beta} = \frac{(p-1)ne^2\nu_p}{4\sqrt{3}c(\gamma_1^{1-p}-\gamma_2^{1-p}) }\left(\frac{\nu}{\nu_p}\right)^{-\frac{p-1}{2}}\int_{x_1}^{x_2} dx x^{\frac{p-3}{2}} H^{\alpha \beta} (\nu, \theta_B),
\end{equation}

\noindent where $\nu_p = \nu_c / \gamma^2$. Then the three emissivities can be written:

\begin{eqnarray}\label{plemis}
j_I &=& \frac{ne^2(p-1)\nu_p}{2\sqrt{3}c (\gamma_1^{1-p}-\gamma_2^{1-p})}\left(\frac{\nu}{\nu_p}\right)^{-\frac{p-1}{2}}[G_I(x_1)-G_I(x_2)],\\
j_Q &=&  \frac{ne^2(p-1)\nu_p}{2\sqrt{3}c (\gamma_1^{1-p}-\gamma_2^{1-p})}\left(\frac{\nu}{\nu_p}\right)^{-\frac{p-1}{2}}[G_Q(x_1)-G_Q(x_2)],\\
j_V &=& \frac{2ne^2(p-1)\nu_p \cot{\theta_B}}{3\sqrt{3}c (\gamma_1^{1-p}-\gamma_2^{1-p})}\left(\frac{\nu}{\nu_p}\right)^{-\frac{p}{2}}[G_V(x_1)-G_V(x_2)],\\
\end{eqnarray}

\noindent where the power law synchrotron integrals are,

\begin{eqnarray}
G_I(x)&=&\int_x^\infty dz z^{\frac{p-3}{2}} F(z),\\
G_Q(x)&=&\int_x^\infty dz z^{\frac{p-3}{2}} G(z),\\
G_V(x)&=&\int_x^\infty dz z^{\frac{p}{2}-1} H(z).
\end{eqnarray}

\noindent In many prior studies \citep{leggwestfold1968,blumenthalgould1970,melrose1971,jonesodell1977}
the integrals are performed analytically for the frequency range
$\gamma_1^2 \nu_p \ll \nu \ll \gamma_2^2 \nu_p$ where the limits of integration,
$x_{1,2} = \nu/(\gamma_{1,2}^2 \nu_p)$ can be extended to $0$ and $\infty$.

For the primary non-thermal source of interest, M87, 

\begin{equation}
\gamma_1^2 \nu_p \sim 2\times10^{11} \left(\frac{B}{10 G}\right)
\left(\frac{\gamma_1}{100}\right)^2 \text{Hz},
\end{equation}

\noindent uncomfortably close to frequencies $\simeq 230$ GHz of interest for mm-VLBI
\citep{doelemanetal2012,akiyamaetal2015} for $\gamma_1 \gtrsim 30$. We
then keep the finite limits of integration and numerically tabulate
the integrals $G_I$, $G_Q$ and $G_V$ as functions of $x$ for select
values of $p = 3.0, 3.5, 7.0$ currently. This procedure can be sped up
significantly using the relation \citep{westfold1959}, 

\begin{eqnarray}
L(x; s, \alpha) &\equiv& \int_x^\infty d\xi \xi^{s-1} \int_\xi^\infty dy K_\alpha(y)\nonumber\\
&=&\frac{\alpha+s}{s}\int_x^\infty d\xi \xi^{s-1} K_\alpha(\xi)-\frac{x^s}{s}\left[\int_x^\infty dy K_{\alpha+1}(y)-K_\alpha(x)\right]
\end{eqnarray}

\noindent to reduce the double integrals to single integrals. The
results agree with those in Legg \& Westfold equation (33) after using a
recurrence relation,  

\begin{equation}
2K_{\alpha}'(x)=-(K_{\alpha+1}+K_{\alpha-1}),
\end{equation}

\noindent and noting that $K_{-\alpha}(x)=K_{\alpha}(x)$ to transform the Bessel functions in $H(x)$. 

To check against approximate formulae elsewhere, we extend the limits of integration to $0$ and $\infty$ and use,

\begin{eqnarray}
I(s,\alpha)&\equiv&\int_0^\infty dx x^s K_\alpha(x)=2^{s-1}\Gamma\left(\frac{s+\alpha+1}{2}\right)\Gamma\left(\frac{s-\alpha+1}{2}\right)\\
J(s,\alpha)&\equiv& L(0;s+1,\alpha-1)=\frac{\alpha+s}{s+1}I(s,\alpha-1)\nonumber\\
&=&\frac{\alpha+s}{s+1}2^{s-1}\Gamma\left(\frac{s+\alpha}{2}\right)\Gamma\left(\frac{s-\alpha}{2}+1\right)
\end{eqnarray}

\noindent to find the approximate forms
$G_{I,Q,V}(0)-G_{I,Q,V}(\infty)$:

\begin{eqnarray}
G_I(0)&=&J\left(\frac{p-1}{2},\frac{5}{3}\right)=\frac{2^{\frac{p-3}{2}}(p+7/3)}{p+1}\Gamma\left(\frac{p}{4}+\frac{7}{12}\right)\Gamma\left(\frac{p}{4}-\frac{1}{12}\right),\\
G_Q(0)&=&I\left(\frac{p-1}{2},\frac{2}{3}\right)=\frac{p+1}{p+7/3} G_I(0),\\
G_V(0)&=&J\left(\frac{p}{2}-1,\frac{1}{3}\right)+I\left(\frac{p}{2},\frac{1}{3}\right)\nonumber\\
&=&\frac{2^{\frac{p}{2}-1}(p+2)}{p}\Gamma\left(\frac{p}{4}+\frac{1}{3}\right)\Gamma\left(\frac{p}{4}+\frac{2}{3}\right),\\
G_{I,Q,V}(\infty)&=&0
\end{eqnarray}

\noindent leading to the approximate emissivities:

\begin{eqnarray}
j_I^0&=&\frac{ne^2(p-1)\nu_p}{2\sqrt{3}c(\gamma_1^{1-p}-\gamma_2^{1-p})}\left(\frac{\nu}{\nu_p}\right)^{-\frac{p-1}{2}}2^{\frac{p-3}{2}}\frac{p+7/3}{p+1}\Gamma\left(\frac{p}{4}+\frac{7}{12}\right)\Gamma\left(\frac{p}{4}-\frac{1}{12}\right),\\
j_Q^0&=&\frac{p+1}{p+7/3}j_I^0,\\
j_V^0&=&\frac{2ne^2(p-1)\nu_p\cot{\theta_B}}{3\sqrt{3}c(\gamma_1^{1-p}-\gamma_2^{1-p})}\left(\frac{\nu}{\nu_p}\right)^{-\frac{p}{2}}2^{\frac{p}{2}-1}\frac{p+2}{p}\Gamma\left(\frac{p}{4}+\frac{1}{3}\right)\Gamma\left(\frac{p}{4}+\frac{2}{3}\right).
\end{eqnarray}

\noindent These results agree with those of several authors. 

In the case of non-thermal emission, the absorption coefficient cannot be simply related to the emissivity using Kirchoff's Law, and instead we use \citep{melrose1980}:

\begin{equation}
\alpha^{\alpha\beta}=-\frac{c}{m\nu^2}\int_0^\infty d\gamma \gamma^2 \frac{d}{d\gamma}\left[\frac{N(\gamma)}{\gamma^2}\right] \eta^{\alpha\beta} (\gamma,\nu,\theta_B).
\end{equation}

\noindent The derivation is analogous to that for the emissivity, and the results are:

\begin{eqnarray}\label{plabs}
\alpha_I &=& \frac{ne^2(p-1)(p+2)}{4\sqrt{3}mc\nu_p(\gamma_1^{1-p}-\gamma_2^{1-p})}\left(\frac{\nu}{\nu_p}\right)^{-\frac{p}{2}-2}\left[Ga_I(x_1)-Ga_I(x_2)\right],\\
\alpha_Q &=& \frac{ne^2(p-1)(p+2)}{4\sqrt{3}mc\nu_p(\gamma_1^{1-p}-\gamma_2^{1-p})}\left(\frac{\nu}{\nu_p}\right)^{-\frac{p}{2}-2}\left[Ga_Q(x_1)-Ga_Q(x_2)\right],\\
\alpha_V &=& \frac{ne^2(p-1)(p+2)\cot{\theta_B}}{3\sqrt{3}mc\nu_p(\gamma_1^{1-p}-\gamma_2^{1-p})}\left(\frac{\nu}{\nu_p}\right)^{-\frac{p+5}{2}}\left[Ga_V(x_1)-Ga_V(x_2)\right]
\end{eqnarray}

\noindent where the power law absorption integrals are,

\begin{eqnarray}
Ga_I(x)&=&\int_x^\infty dz z^{\frac{p}{2}-1} F(z),\\
Ga_Q(x)&=&\int_x^\infty dz z^{\frac{p}{2}-1} G(z),\\
Ga_V(x)&=&\int_x^\infty dz z^{\frac{p-1}{2}} H(z).
\end{eqnarray}

\noindent Again extending the limits of integration, we find agreement with approximate formulae in the literature:

\begin{eqnarray}
Ga_I(0)&=&\frac{p+10/3}{p+1}2^{\frac{p}{2}-1}\Gamma\left(\frac{p}{4}+\frac{5}{6}\right)\Gamma\left(\frac{p}{4}+\frac{1}{6}\right),\\
Ga_Q(0)&=&\frac{p+2}{p+10/3}Ga_I(0)\\
Ga_V(0)&=&\frac{p+3}{p+1}2^{\frac{p-1}{2}}\Gamma\left(\frac{p}{4}+\frac{7}{12}\right)\Gamma\left(\frac{p}{4}+\frac{11}{12}\right),\\
Ga_{I,Q,V}(\infty)&=&0.
\end{eqnarray}

\noindent Figure \ref{plcompare} compares numerical integration of the 
formulae in Equations (\ref{plemis}) and (\ref{plabs}) with the forms from
the literature. 

\section{Faraday coefficients for power law and thermal distributions
  of electrons}
\label{sec:farad-coeff-power}

\begin{figure}
\begin{tabular}{ll}
\includegraphics[scale=0.58]{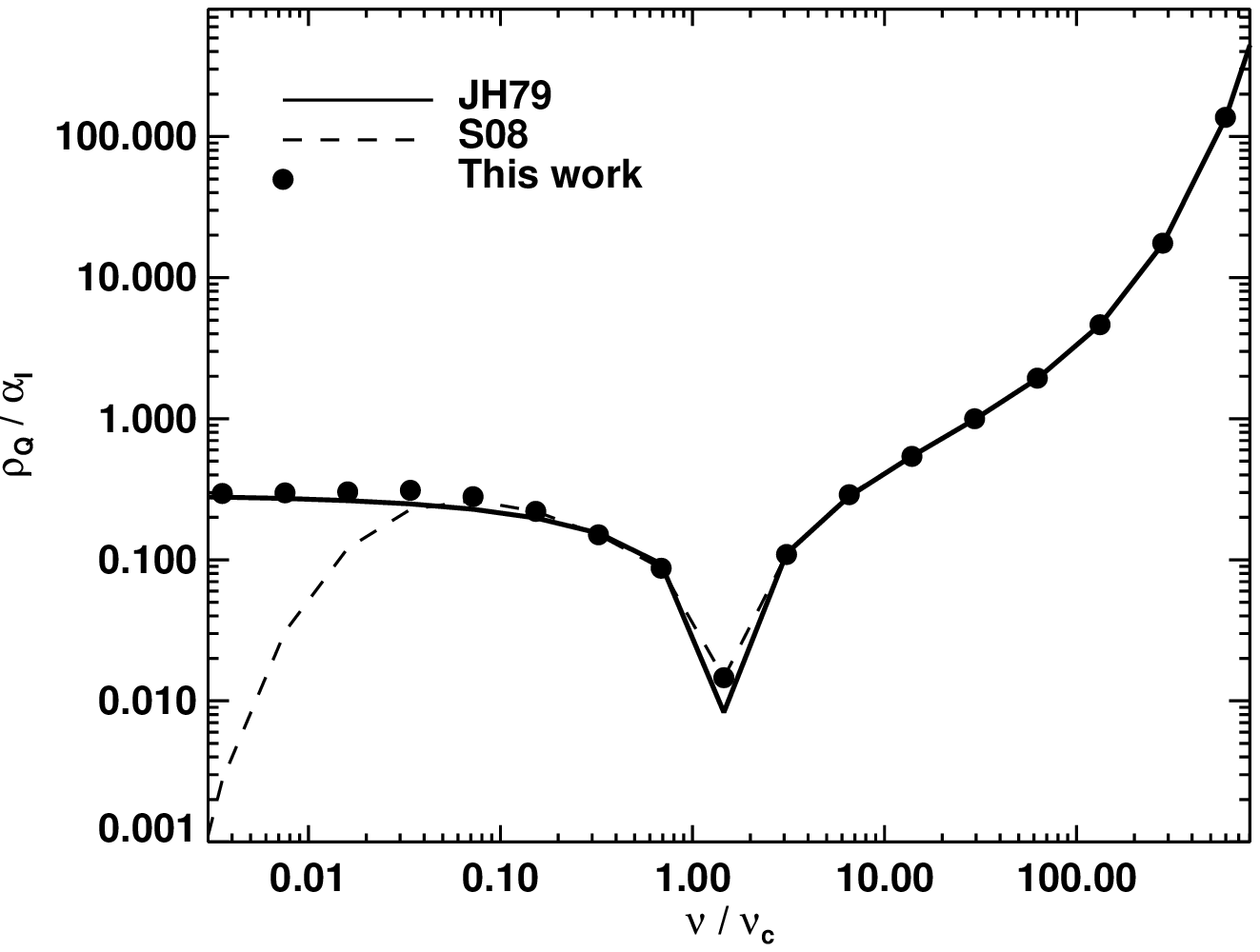}&
\includegraphics[scale=0.58]{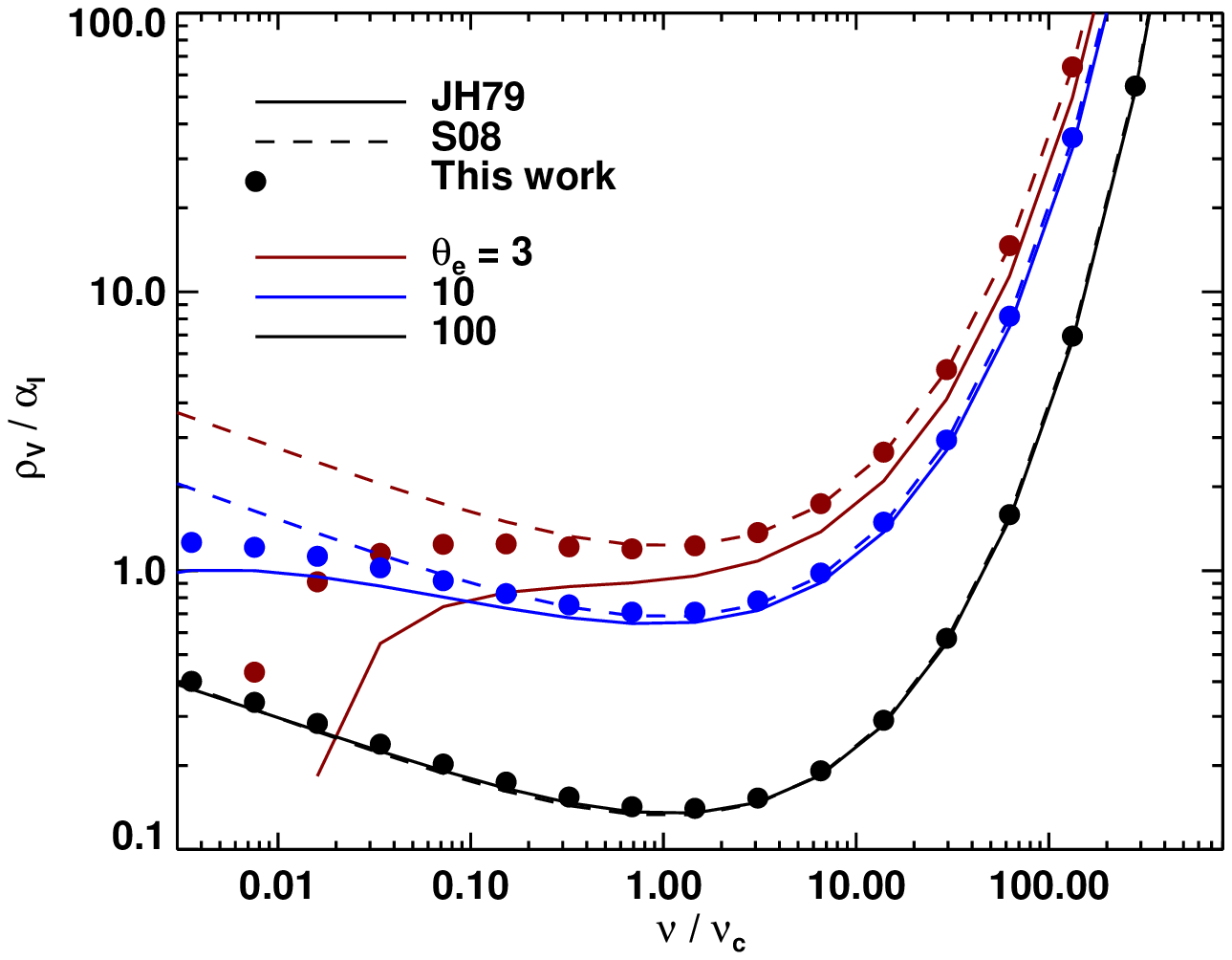}\\
\end{tabular}
\caption{\label{fig:rotcoef}\emph{Left:} Ratio of thermal Faraday conversion
  coefficient to total absorption as a function of $\nu/\nu_c$ at
  $\theta_e=100$ calculated from numerical integration of the
  expression in \citet{joneshardee1979} (solid line,
  equation \eqref{joneshardeerho}), the fitting function approach introduced by
  \citet{shcherbakov2008} (dashed line, equation \eqref{shcherbakovrho}), and in this
  work (black dots, equation \eqref{ourrhoq}). \emph{Right:} Same as the
  left panel but for Faraday rotation at 3 different temperatures. In
  both cases all results agree at high frequencies and
  temperatures. Our fitting functions use the 
  temperature dependence from \citet{shcherbakov2008} along with the
  low-frequency limits from \citet{joneshardee1979}.}
\end{figure}

As well as the emission and absorption coefficients calculated above
(Appendix \ref{cha:polar-synchr-emiss}), the 
coefficients $\rho_V$ and $\rho_Q$ affect the generation and transfer
of polarisation in a magnetised plasma. We use approximate expressions
from the literature for these coefficients which as above are modified to i) be
fast to evaluate and ii) have the correct asymptotic limits. 

\subsection{Power law distribution}

In the case of a power law distribution, we use the expressions from
\citet{jonesodell1977} Appendix C, written in our notation:

\begin{eqnarray}
\rho_Q &=& -\rho_\perp \left(\frac{\nu_B}{\nu}\right)^3 \gamma_{\rm
  min}^{2-p}\left[\left(1-\left(\frac{\nu_{\rm
        min}}{\nu}\right)^{p/2-1}\right)\left(p/2-1\right)^{-1}\right],\\
\rho_V &=&2 \frac{p+2}{p+1} \rho_\perp\left(\frac{\nu_B}{\nu}\right)^2 \gamma_{\rm min}^{-(p+1)}\ln
\gamma_{\rm min} \cot{\theta_B},\\
\rho_\perp&=&\frac{n e^2}{m c \nu_B} (p-1)  \left[\gamma_{\rm min}^{1-p}-\gamma_{\rm max}^{1-p}\right]^{-1}.
\end{eqnarray}

\noindent More accurate expressions \citep{huangshcherbakov2011}
require integration over the distribution function and for this reason 
are slow to evaluate. These approximate forms are relatively accurate
for $\gamma_{\rm min} \lesssim 10^2$ \citep[left panel of Fig. 6
in][]{huangshcherbakov2011}. In the example semi-analytic jet problem
above, Faraday rotation and conversion are negligible
\citep{broderickloeb2009}. Nonetheless, it should be possible to find
accurate fitting functions for these coefficients, which would be
consistent with our approach for the other coefficients. 

\subsection{Thermal distribution}
\label{sec:faraday-thermal-distribution}

Faraday coefficients for thermal distributions of electrons have been
calculated in limits of either high frequency $\nu/\nu_c \gg 1$
\citep[e.g.,][]{melrose1997}, at high temperatures $\theta_e \gg 1$, or
both. In particular, \citet{shcherbakov2008} provided approximate
fitting functions for $\rho_V$ and $\rho_Q$ over a wide temperature
range $\theta_e \gtrsim 1$ with high accuracy for $\nu / \nu_c \gtrsim
10^{-1}$ (their equations 25, 26, 33, but in our notation):

\begin{eqnarray}\label{shcherbakovrho}
\rho_Q &=& \frac{n e^2 \nu_B^2 \sin{\theta_B}^2}{m c \nu^3} f(X)
\left[\frac{K_1 (\theta_e^{-1})}{K_2(\theta_e^{-1})}+6 \theta_e\right]\\
\rho_V &=& \frac{2 n e^2 \nu_B}{m c \nu^2}
\frac{K_0(\theta_e^{-1})}{K_2 (\theta_e^{-1})} \cos{\theta_B} g(X),
\end{eqnarray}

\noindent where

\begin{eqnarray}
f(X) &=& 2.011 \exp\left(-\frac{X^{1.035}}{4.7}\right)-\cos\left(\frac{X}{2}\right)\exp\left(-\frac{X^{1/2}}{2.73}\right)-0.011\exp\left(-\frac{X}{47.2}\right)\\
g(X) &=& 1-0.11 \ln \left(1+0.035X\right)\\
X &=& \left(\frac{3}{2\sqrt{2}}10^{-3} \frac{\nu}{\nu_c} \right)^{-1/2},
\end{eqnarray}

\noindent and their parameter $X$ is a function of $\nu /
\nu_c$. In the high-frequency limit $\nu/\nu_c \gg 1$, both functions
asymptotically reach unity. The cosine term in $f(X)$ is used to fit
the sign change in $\rho_Q$ near $\nu/\nu_c \simeq 1.5$.

While the high frequency and high temperature limit is most
relevant for our applications of interest, e.g. modeling the submm
emission from Sgr A*, ideally we
would have expressions that are correct in both asymptotic limits. For
this reason we modify the expressions in \citet{shcherbakov2008}, by
comparing them with the expressions for the high temperature 
synchrotron limit given in \citet{joneshardee1979} (their equations 3-4, in
our notation):

\begin{eqnarray}\label{joneshardeerho}
\rho_Q &=& \frac{\pi n e^2 (\nu / \nu_c)^{-2/3}}{2^{4/3} 3^{2/3} m c
  \theta_e^3 \nu}  J_4 (\nu / \nu_c), \\
\rho_V &=& \frac{n e^2 \nu_B \cos{\theta_B}}{m c \nu^2 \theta_e^2} J_5
(\nu / \nu_c, \theta_e, \theta_B),
\end{eqnarray}

\noindent where

\begin{eqnarray}
J_4 (z) &=& \int_0^\infty dy y^{4/3} Gi'(q) e^{-y},\\
J_5 (z,\theta_e,\theta_B) &=& \int_0^\infty dy \left[\ln (y \theta_e) +
    \sin{\theta_B} \ln 2 + 1/3 \left[q \pi Gi(q)-1\right]+1/2\int_q^\infty dx \left(\pi
    Gi(x) - 1/x\right)\right] e^{-y},
\end{eqnarray}

\noindent and where $q \equiv (3/2 z/y^2)^{2/3}$ and $Gi(x)$ is defined in terms of Airy functions \citep{abramowitzstegun1970}.

In the high-frequency, high temperature limit where $\nu / \nu_c \gg 1$,
$\theta_e \gg 1$, and where we can replace the Bessel functions by
their asymptotic limits and $J_4 \rightarrow -24 / \pi (3/2 z)^{-4/3}$,
$J_5 \rightarrow \ln \theta_e$, these expressions
agree with the above results from \citet{shcherbakov2008}. From
numerically integrating $J_4 (z)$ and $J_5 (z)$, we also find good
agreement between the two sets of coefficients over their reported 
ranges of validity. We adapt the fitting function $f(X)$
in \citet{shcherbakov2008} to use the asymptotic limit of $\rho_Q$ at
small $\nu/\nu_c$, where $J_4 (z) \rightarrow \frac{4}{9 \times 3^{5/6}}$:

\begin{eqnarray}\label{ourrhoq}
f_m(X) = f(X) + \left[0.011 \exp{\left(-\frac{X}{47.2}\right)}-2^{-1/3}/3^{23/6}
  10^4 \pi X^{-8/3}\right] 1/2 \left[1+\tanh{\left(10 \ln {x/120}\right)}\right].
\end{eqnarray}

\noindent The added term imposes the correct
asymptotic limit at large $X$ (small $\nu/\nu_c$) and for this reason
maintains good accuracy over all $\nu/\nu_c$. 

For $\rho_V$, the term $J5(\nu/\nu_c)$ from
\citet{joneshardee1979} separates into a sum of terms which depend on
temperature and $\theta_B$, and those which only depend on
$\nu/\nu_c$. This suggests that it would be better to use the
factor $g(X)$ in \citet{shcherbakov2008} as a difference from the
high-frequency limit rather than a multiplication:

\begin{equation}\label{ourrhov}
\rho_V = \frac{2 n e^2 \nu_B \cos{\theta_B}}{m c \nu^2}\frac{K_0(\theta_e^{-1}) - \Delta J_5(X)}{K_2 (\theta_e^{-1})},
\end{equation}

\noindent where we define our correction factor $\Delta J_5(X)$ in a
similar spirit to that of \citet{shcherbakov2008}, but modified for
higher accuracy:

\begin{equation}
\Delta J_5(X) = 0.4379 \ln(1+0.001858 X^{1.503}).
\end{equation}

Figure \ref{fig:rotcoef} compares numerical integration of equation \eqref{joneshardeerho}
\citep{joneshardee1979} with the fitting functions from equation \eqref{shcherbakovrho}
\citep{shcherbakov2008} and our modified forms. Our approximate 
versions are fast to compute, while maintaining accuracy over all
$\nu/\nu_c$. 

We have compared polarised spectra and maps from \textsc{HARM} models of Sgr A* computed using
the prescriptions from \citet{shcherbakov2008},
\citet{joneshardee1979}, and the high temperature, high frequency
limit. As expected, all expressions are in excellent agreement when
$\nu / \nu_c \gg 1$, in this case for $\nu \gtrsim 10^{12}$ Hz. Below
that, our modified expressions and those from \citet{shcherbakov2008}
are in good agreement, while the degree of circular polarisation can
differ between our results and the high-frequency limit. As described
in the main text, models of Sgr A* in the submm have self-absorption
optical depth $\tau_I \gtrsim 1$, $\nu / \nu_c \sim 10-100$, and
$\theta_e \sim 10$, so that the Faraday optical depths can be very
large and have an important effect on the resulting polarisation maps
and spectra (e.g. Figure \ref{harmpol}). Since this result is for the
high-frequency, relativistic limit, it does not depend on the fitting
function used.

\section{Analytic solutions to the polarised radiative transfer
  equations}
\label{sec:analyt-solut-polar}

This appendix provides the analytic solutions to the polarised
radiative transfer equations used for testing
different integration methods for \textsc{grtrans} in \S \ref{sec:integration-tests}. In all cases, the boundary condition used is that
the initial intensity is zero for each Stokes parameter.

The first case considered is pure emission and absorption in stokes
$I$ and $Q$, in which case equation \eqref{transfer} becomes:

\begin{equation}\label{transferabs}
\frac{d}{ds}
\left(\begin{array}{c}  I \\  Q \\ \end{array}\right)
=
\left(\begin{array}{c}
  j_I \\  j_Q \\
\end{array}\right)-
\left(%
\begin{array}{cc}
  \alpha_I & \alpha_Q \\
  \alpha_Q & \alpha_I \\
\end{array}%
\right)
\left(\begin{array}{c}  I \\  Q \\ \end{array}\right),
\end{equation}

\noindent whose solution is,

\begin{eqnarray}\label{eq:2}
I(s) &=& \frac{1}{\alpha (\alpha_I - \alpha_Q)} \left\{(j_I \alpha_I -
    j_Q \alpha_Q) \left[1 - \frac{e^{-\alpha s}}{2}(1 + e^{2 \alpha_Q
      s})\right] + (j_I \alpha_Q - j_Q \alpha_I) \frac{e^{-\alpha s}}{2}
    (1-e^{2 \alpha_Q s})\right\},\\
Q(s) &=& \frac{1}{\alpha (\alpha_I - \alpha_Q)} \left\{(j_Q \alpha_I - j_I \alpha_Q)\left[1-\frac{e^{-\alpha
    s}}{2}\left(1+e^{2\alpha_Q s}\right)\right]  + (j_Q \alpha_Q - j_I \alpha_I) \frac{e^{-\alpha s}}{2} \left(1 - e^{2 \alpha_Q s}
  \right)\right\},
\end{eqnarray}

\noindent where $\alpha \equiv \alpha_I + \alpha_Q$.  When $\alpha_Q =
0$, the second group of terms in each equation vanishes while the first reduces to the usual formal
solution of the radiative transfer equation,
e.g. when stokes $I$ and $Q$ are not coupled. Examples of the solution are shown in Figure \ref{testabs}.

The second case of interest is pure polarised emission in Stokes
$(Q,U,V)$ along with Faraday rotation and conversion
($\rho_{Q,V}$). Here the polarised radiative transfer equation is,

\begin{equation}\label{transferfar}
\frac{d}{ds}
\left(\begin{array}{c}  Q \\  U \\  V \\\end{array}\right)
=
\left(\begin{array}{c}
 j_Q \\  j_U \\  j_V
\end{array}\right)-
\left(%
\begin{array}{cccc}
  0 & \rho_V & 0 \\
  -\rho_V & 0 & \rho_Q \\
 0 & -\rho_Q & 0 \\
\end{array}%
\right)
\left(\begin{array}{c}  Q \\  U \\  V \\\end{array}\right),
\end{equation}

\noindent where we have set $\rho_U = 0$ as is commonly chosen for the
Stokes basis for synchrotron radiation. From this equation it is
apparent that $\rho_V$ is responsible for changing the linear
polarisation direction (mixing stokes $Q$ and $U$, Faraday rotation)
while $\rho_Q$ converts between linear and circular polarisation
(mixing stokes $U$ and $V$, Faraday conversion). The solution is,

\begin{eqnarray}\label{eq:1}
Q(s) &=&  \frac{\rho_Q}{\rho^2} (j_Q \rho_Q + j_V \rho_V) s -
\frac{\rho_V}{\rho^3} (j_V \rho_Q - j_Q \rho_V) \sin{\rho s} -
\frac{j_U \rho_V}{\rho^2} (1 - \cos{\rho s})\\
U(s) &=& \frac{j_Q \rho_V - j_V \rho_Q}{\rho^2} (1 - \cos{\rho s}) +
\frac{j_U}{\rho} \sin{\rho s},\\
V(s) &=& \frac{\rho_V}{\rho^2} (j_Q \rho_Q + j_V \rho_V) s -
\frac{\rho_Q}{\rho^3} (j_Q \rho_V - j_V \rho_Q) \sin{\rho s} +
\frac{j_U \rho_Q}{\rho^2} (1 - \cos{\rho s}),\\
\end{eqnarray}

\noindent where $\rho \equiv \sqrt{\rho_Q^2+\rho_V^2}$. These
solutions for a sample case are plotted in Figure \ref{testfar}.

In the case of only Faraday rotation or conversion ($\rho_Q = 0$ for Stokes $Q$ or
$\rho_V = 0$ for Stokes $V$), the solution is purely
oscillatory with the maximum linearly polarised
intensity restricted to be $\sim j / \rho$ independent of
the total intensity or path length, despite the fact that there is no
absorption. Since in this optically thin limit the total intensity
grows as $j_I s$, the fractional polarisation decreases as $1 /
s$. When both Faraday rotation and conversion are present, the Stokes
$Q$ and $V$ acquire terms which linearly increase with $s$, while the
oscillatory terms still have maximum values that are independent of
$s$. This means that in the limit of large Faraday optical depth
(large $s$), the fractional polarisation approaches a constant value
$Q, V / I = \rho_{Q,V} (j_Q \rho_Q + j_V \rho_V) / j_I \rho^2$ instead
of decreasing as $1 / s$ in the pure Faraday rotation or conversion
case. Since for cases of interest $\rho_V > \rho_Q$, circular
polarisation becomes dominant over linear polarisation in the limit of
large Faraday optical depth.

This is a different limit than an initial polarised intensity
travel through a magnetised medium where Faraday rotation occurs. In
that case, the intensity also oscillates between Stokes $Q$ and $U$,
but with a constant polarised intensity. The fractional polarisation
only decreases when the Faraday rotation is
instead occurring in the region where the polarised emission is being
produced.

\section{Closed form expression for $\mathbf{O}(s,s')$}
\label{deglinnocentio}

\citet{deglinnocenti1985} found a closed form solution for the matrix
operator $\mathbf{O}(s,s')$, defined by

\begin{eqnarray}
\frac{d}{ds} \mathbf{O}(s,s') = -\mathbf{K}(s) \mathbf{O}(s,s'), \hspace{12pt}\mathbf{O}(s,s) = 1,
\end{eqnarray}

\noindent which describes the transfer of the Stokes parameters from position
$s$ to $s'$ in the absence of emission,
which is valid under limited conditions including when the absorption 
matrix $\mathbf{K}$ is constant over the interval. We reproduce the solution
here in our notation: 

\begin{equation}
\mathbf{O}(s,s') = \exp(-\alpha_I \Delta s) \left\{\left[\cosh{\left(\Lambda_1
        \Delta s\right)}+\cos{\left(\Lambda_2 \Delta s\right)}\right] \mathbf{M_1} / 2 -
    \sin{\left(\Lambda_2 \Delta s\right)} \mathbf{M_2} - \sinh{\left(\Lambda_1
      \Delta s\right)} \mathbf{M_3}+\left[\cosh{\left(\Lambda_1 \Delta
        s\right)}-\cos{\left(\Lambda_2 \Delta s\right)}\right]\mathbf{M_4} / 2\right\},
\end{equation}

\noindent where

\begin{eqnarray}
\mathbf{M_1} &=& \mathbf{1}\\
\mathbf{M_2} &=&\frac{1}{\Theta}\left(%
\begin{array}{cccc}
  0 & \Lambda_2 \alpha_Q - \sigma \Lambda_1 \rho_Q & \Lambda_2
  \alpha_U - \sigma \Lambda_1 \rho_U & \Lambda_2 \alpha_V - \sigma
  \Lambda_1 \rho_V \\
  \Lambda_2 \alpha_Q - \sigma \Lambda_1 \rho_Q & 0 &\sigma \Lambda_1
  \alpha_V + \Lambda_2 \rho_V & -\sigma \Lambda_1 \alpha_U - \Lambda_2
  \rho_U\\
 \Lambda_2 \alpha_U - \sigma \Lambda_1 \rho_U & -\sigma \Lambda_1
 \alpha_V - \Lambda_2 \rho_V & 0 & \sigma \Lambda_1 \alpha_Q +
 \Lambda_2 \rho_Q \\
 \Lambda_2 \alpha_V - \sigma \Lambda_1 \rho_V & \sigma \Lambda_1
 \alpha_U + \Lambda_2 \rho_U & -\sigma \Lambda_1 \alpha_Q - \Lambda_2 \rho_Q & 0 \\
\end{array}%
\right)\\
\mathbf{M_3} &=&\frac{1}{\Theta}\left(%
\begin{array}{cccc}
  0 & \Lambda_1 \alpha_Q + \sigma \Lambda_2 \rho_Q & \Lambda_1
  \alpha_U + \sigma \Lambda_2 \rho_Q & \Lambda_1 \alpha_V + \sigma
  \Lambda_2 \rho_V \\
  \Lambda_1 \alpha_Q + \sigma \Lambda_2 \rho_Q & 0 & -\sigma \Lambda_2
  \alpha_V + \Lambda_1 \rho_V & \sigma \Lambda_2 \alpha_U - \Lambda_1 \rho_U \\
   \Lambda_1 \alpha_U + \sigma \Lambda_2 \rho_U & \sigma \Lambda_2
   \alpha_V - \Lambda_1 \rho_V & 0 & -\sigma \Lambda_2 \alpha_Q +
   \Lambda_1 \rho_Q \\
   \Lambda_1 \alpha_V + \sigma \Lambda_2 \rho_V & -\sigma \Lambda_2
   \alpha_U + \Lambda_1 \rho_U & \sigma \Lambda_2 \alpha_Q - \Lambda_1
   \rho_Q & 0 \\
\end{array}%
\right)\\
\mathbf{M_4} &=&\frac{2}{\Theta}\left(%
\begin{array}{cccc}
  \left(\alpha^2 + \rho^2\right)/2 & \alpha_V \rho_U - \alpha_U \rho_V
  & \alpha_Q \rho_V - \alpha_V \rho_Q & \alpha_U \rho_Q - \alpha_Q \rho_U \\
  \alpha_U \rho_V - \alpha_V \rho_U & \alpha_Q^2+\rho_Q^2 -
  \left(\alpha^2 + \rho^2 \right)/2 & \alpha_Q \alpha_U + \rho_Q
  \rho_U & \alpha_V \alpha_Q + \rho_V \rho_Q \\
 \alpha_V \rho_Q - \alpha_Q \rho_V &\alpha_Q \alpha_U + \rho_Q \rho_U & \alpha_U^2+\rho_U^2 -
  \left(\alpha^2 + \rho^2 \right)/2 & \alpha_U \alpha_V + \rho_U \rho_V \\
 \alpha_Q \rho_U - \alpha_U \rho_Q & \alpha_V \alpha_Q + \rho_V \rho_Q
 & \alpha_U \alpha_V + \rho_U \rho_V & \alpha_V^2+\rho_V^2 -
  \left(\alpha^2 + \rho^2 \right)/2 \\
\end{array}%
\right)\\
\end{eqnarray}

\noindent and 

\begin{eqnarray}
\Theta &=& 2 \left[\left(\alpha^2 - \rho^2\right)^2/4 + \left(\mathbf{\alpha} \cdot \mathbf{\rho}\right)^2\right]^{1/2},\\
\Lambda_{1,2} &=& \left\{\left[\left(\alpha^2-\rho^2\right)^2/4 +
    \left(\mathbf{\alpha} \cdot \mathbf{\rho}\right)^2\right]^{1/2} \pm \left(\alpha^2 - \rho^2\right)/2\right\}^{1/2},\\ 
\sigma &=& \rm sign \left(\mathbf{\alpha} \cdot
  \mathbf{\rho}\right),\\
\mathbf{\alpha} \cdot \mathbf{\rho} &=& \alpha_Q \rho_Q + \alpha_U \rho_U + \alpha_V
\rho_V,\\
\rho^2 &=& \rho_Q^2 + \rho_U^2+\rho_V^2,\\ 
\alpha^2 &=& \alpha_Q^2 + \alpha_U^2 + \alpha_V^2.
\end{eqnarray}

%%%%%%%%%%%%%%%%%%%%%%%%%

% Don't change these lines
\bsp	% typesetting comment
\label{lastpage}
\end{document}